\documentclass[a4paper,11pt]{article}
\pdfoutput=1 

\usepackage{jheppub} 

\usepackage[T1]{fontenc} 

\usepackage{amsmath}
\usepackage{amssymb}
\usepackage{graphicx}
\usepackage{bm}
\usepackage{epsfig}
\usepackage{amsfonts}
\usepackage{color}
\usepackage{mathtools}

\title{A first look at Bottomonium melting via a stochastic potential}

\author[a]{Alexander Rothkopf}

\affiliation[a]{Albert Einstein Center for Fundamental Physics,  University of Bern, \\ Sidlerstr.5, 3012 Bern, Switzerland}

\emailAdd{rothkopf@itp.unibe.ch}

\abstract{We investigate the phenomenon of Bottomonium melting in a thermal quark-gluon plasma using three-dimensional stochastic simulations based on the concept of open-quantum systems. In this non-relativistic framework, introduced in \cite{Akamatsu:2011se}, which makes close contact to the potentials derived in effective field theory, the $b\bar{b}$ system evolves unitarily under the incessant kicks by the constituents of the surrounding heat bath. In particular thermal fluctuations and the presence of a complex potential in the EFT are naturally related.
An intricate interplay between state mixing and thermal excitations emerges as we show how non-thermal initial conditions of Bottomonium states evolve over time. We emphasize that the dynamics of these states gives us access to information beyond what is encoded in the thermal Bottomonium spectral functions. Assumptions underlying our approach and their limitations, as well as the refinements necessary to connect to experimental measurements under more realistic conditions are discussed.}

\begin{document} 

\maketitle

\flushbottom

\section{Bottomonium as QGP thermometer}

Elucidating the properties of strongly interacting matter shortly after the Big-Bang is the central reason behind scrutinizing the collisions of heavy-ions \cite{Muller:2013dea} at relativistic energies at the RHIC and the LHC collider facilities. Under the extremely large energy densities present in the very early universe as well as in the center of these man made collisions, the fundamental matter constituents, quarks and gluons, are not confined within hadrons but instead form a quark-gluon plasma (QGP) \cite{Yagi:2005yb,Sarkar:2010zza}.

Measurements of e.g. the elliptic flow of light hadrons at RHIC \cite{Afanasiev:2009wq,Adamczyk:2011aa} already helped to establish that a very low shear viscosity to entropy ratio $\eta/s \gtrsim 1/4\pi$ \cite{Song:2010mg} is a prime characteristic of this high temperature phase of quantum chromo dynamics (QCD), just above the deconfinement transition. The precise determination of such bulk properties of the fireball created in the collision center, ideally crosschecked by multiple independent methods, is steadily progressing but remains a challenging endeavor. 

Another phenomenologically interesting property is the temperature reached in heavy-ion collisions. On the one hand electromagnetic probes, such as thermal photons emitted from the QGP have been utilized to obtain estimates both for $T_{\rm RHIC}(\sqrt{s_{NN}}=200{\rm GeV})=221\pm19 \rm MeV$ and $T_{\rm LHC}(\sqrt{s_{NN}}=2.76{\rm TeV})=304\pm51\rm MeV$ \cite{Adare:2008ab,Wilde:2012wc}, which lie well above the deconfinement crossover transition $T\sim150\rm MeV$ determined in numerical lattice QCD studies \cite{Aoki:2006we,Borsanyi:2010bp,Bazavov:2011nk}. In this study on the other hand we focus solely on strongly interacting objects and how they can be used to infer medium properties. Describing their evolution hence requires us to deal with the real-time dynamics of quarks and gluons at finite temperature. 

In general, to measure the temperature $T$ of a physical system, we can resort to either a static or a dynamic strategy. The former, a thermometer in the everyday sense, requires us to bring into contact with the system of interest a test system, which after having equilibrated over time to a common temperature we decouple and inspect regarding its own thermodynamics properties, such as its density, free energy or entropy. The dynamical approach on the other hand observes the not necessarily thermal properties of a test system as it approaches equilibrium with its surroundings, just as the speed of melting a sugar cube in a cup of tea reveals the temperature of the beverage. While on macroscopic scales both approaches are often applicable, the physics at the subatomic realm, in particular within the strongly interacting QGP at the center of a heavy-ion collision, poses two important challenges to choosing an appropriate tactic. 

First we need to find appropriate probes, whose behavior is sensitive to the surrounding medium, while their underlying structure remains largely unaffected. I.e. we require probes whose characteristic internal scales, such as their masses $M$, lie well above those of the medium partons. The more pronounced this separation is, the more straight forward their theoretical description becomes, as there exists a small parameter (dividing the scale of the medium by the scale of the probe), which can be used as an expansion parameter to construct effective approximate descriptions.  

Secondly the system we wish to investigate only exists for a minute amount of time $\tau_{\rm QGP}\sim 5-10 \rm fm$, and has reached local thermal equilibrium only after $\tau_{\rm QGP}^{\to T}\sim 1 \rm fm$. As we require a clear separation of scales for our probes $M\gg T_{\rm QGP}$, we assume here that their thermalization time, i.e. the time it takes for their number density to become Bose-Einstein (or Fermi-Dirac) distributed, to be comparatively long $\tau_{\rm probe}^{\to T}\sim\frac{M}{T_{\rm QGP}}\tau_{\rm QGP}^{\to T}$. Hence we will resort to a dynamic strategy for temperature measurements, as we anticipate that observing the thermal properties of our probes will be difficult within the lifetime of the QGP. 

One popular candidate for an appropriate kind of probe are the bound states of heavy quarks and anti-quarks, so called heavy quarkonium. Here we will work with the Bottomonium family, in particular the vector mesons consisting of a $b$ and $\bar{b}$ quark, with a constituent mass $m_b=4.66\pm0.03\rm GeV$ (1S) \cite{Beringer:1900zz}. 

The clear separation between $m_b\gg T_{\rm LHC}, T_{\rm RHIC}$ underlies the intuitive notion that Bottomonium and its interactions with the surrounding medium can be captured in an effective non-relativistic description. More precisely, one wishes to summarize the in-medium physics of the two-body system in an effective potential $V^{\rm EFT}({\bf r})$ entering a non-relativistic Schroedinger equation. 
The formalism of effective field theory \cite{Brambilla:1999xf,Brambilla:2004jw,Brambilla:2008cx,Rothkopf:2013ria,Ghiglieri:2013iya} adapted to finite temperature over the last decade, allows to put this idea onto a solid theoretical footing. 

On the experimental side, the advent of the LHC has made it possible to produce $b\bar{b}$ abundantly, such that the production of e.g. the S-wave bound states in $\rm p+p$ as well as $\rm Pb+Pb$ has been determined with unprecedented precision \cite{Chatrchyan:2011pe}. While calibration of absolute yields is ongoing \cite{Satz:2013ama,OpenHF}, the measured relative suppression of excited states in the presence of a deconfined medium already provides a first stringent benchmark for theoretical descriptions of in-medium $b\bar{b}$. 

The large mass of e.g. the ground state $\Upsilon$ also means that its formation time from the $b\bar{b}$ pairs created in the initial hard phase of a heavy-ion collision $\tau_{b\bar{b}\to\Upsilon}\sim r_{\Upsilon}=0.14 \rm fm$ \cite{BMHeavyIon} is much shorter than the QGP thermalization time. Of course the separation of initial production and QGP formation is not as pronounced for the excited states, such as $\chi_b$. Nevertheless we will proceed, based on the working hypothesis that fully formed $b\bar{b}$ bound states enter the quark-gluon plasma. 

This scenario differs markedly from the description originally drafted for the case of charmonium, the bound states formed by a $c$ quark and its anti-partner. Due to the smaller charm mass $m_c=1.275\pm0.025\rm GeV$ $(\overline{MS})$ \cite{Beringer:1900zz} and correspondingly larger charmonium radii, it is expected that many $c\bar{c}$ pairs will be produced in the initial stages of the collision, which subsequently enter the QGP without having formed a fully bound state. In the presence of thermal fluctuations it is then argued that the strong interactions are not capable to support the formation of a bound ground state already at temperatures as low as $T\sim 1.2T_C$. Hence the suppression of the measured abundances was proposed as a prime signal of QGP formation. 

This train of thought, first introduced in the seminal paper \cite{Matsui:1986dk} and later refined e.g. in \cite{Digal:2001ue} relies on an investigation of the properties, such as the radii, of the Eigenstates of the in-medium Hamiltonian for the two-body $c\bar{c}$ system. These solutions describe quarkonium formed from $c\bar{c}$ pairs, which are essentially in thermal equilibrium with the surrounding medium, since they propagate in time simply through a phase factor. 

A similar type of information is contained in the thermal spectral functions $\rho(\omega)$ calculated using the in-medium heavy quarkonium Schroedinger equation \cite{Burnier:2007qm,Mocsy:2007jz,Petreczky:2010tk}, from the T-Matrix approach \cite{Riek:2010fk} or directly from lattice QCD \cite{Umeda:2002vr,Asakawa:2003re,Aarts:2013kaa}. In vacuum the presence of individual delta peaks clearly outlines possible bound states for the $c\bar{c}$ pairs to end up in. At finite temperature, broadening and disappearance of these structures is related to the fact that Debye screening and Landau damping \cite{Beraudo:2007ky} weakens the net attractive forces responsible to bind thermal $q\bar{q}$ pairs into a well defined heavy quarkonium. 

To conclude as to whether the heavy quarkonium bound state will be able to form or not however relies on an inspection by eye. The absence of a clear concept of what constitutes a bound state in this otherwise theoretically solid assessment leads to ambiguities, reflected in different interpretations of the broadening, i.e. threshold enhancement vs. melting, discussed e.g. in \cite{Asakawa:2003re,Mocsy:2007yj}.

In other words, while thermal spectral functions can help us to understand whether fully thermalized $Q\bar{Q}$ pairs will be able to bind into quarkonium, we require different information to approach Bottomonium in a scenario assuming early-time formation and slow thermalization. The $b\bar{b}$ states then traverse the QGP truly as probes and not as part of the medium. Furthermore, as the bound states possess a long lifetime relative to the nuclear scales, their decay into gluons may occur at a point where the QGP has already ceased to exist. Both issues preclude us from making direct comparisons between the calculated Bottomonium spectral functions and the measured dilepton spectra by e.g. the CMS collaboration \cite{Chatrchyan:2011pe}, which are related only if the dileptons are emitted from the annihilations of fully thermalized quark anti-quark pairs within a plasma \cite{Yagi:2005yb,Burnier:2007qm}.

Take the $b\bar{b}$ ground state in the vector channel, Upsilon, for example. When traveling in the QGP, it is not an Eigenstate of the in-medium Hamiltonian and thus will experience mixing with the excited states $\Upsilon'$ and $\Upsilon''$, a phenomenon which has recently been emphasized in \cite{Borghini:2011yq,Borghini:2011ms,Akamatsu:2011se,CasalderreySolana:2012av,Dutta:2012nw}. In addition the thermal fluctuations in the medium will induce additional excitations and de-excitations between the different Bottomonium states along the time evolution. The language of open quantum systems and in particular the concept of stochastic potential will help us to describe this situation in an intuitive fashion.

In section II we briefly review the origin for the Schroedinger equation for heavy quarkonium and the inter-quark potential derived from the underlying field theory QCD via effective field theory methods. The fact that this potential takes on complex values at finite temperature is interpreted in terms of decoherence and we describe how the concept of stochastic potential allows us to set up an approximate description of the Bottomonium real-time dynamics. We emphasize that the thermal fluctuations and the imaginary part of the EFT potential are intimately related in this approach.

For a first and thus necessarily rather crude estimation on the evolution of Bottomonium in the QGP we carry out three dimensional simulations of the stochastic dynamics in section III. We assume a static medium described by the perturbative EFT potential \cite{Laine:2006ns} at $T=2.33T_C$ and observe how individual states as well as non-thermal ensembles of Bottomonium evolve over time. We discuss the dependence of the abundances of heavy quarkonium states on information beyond what is contained in the thermal spectral functions and conclude in section IV with a list of necessary improvements to our method before a quantitative comparison to experiment can be attempted.

\section{From complex to stochastic potential}

\subsection{A Schroedinger equation for heavy quarkonium}

If there exists a separation of scales in a system and we are concerned with the physics only at one of these scales, the framework of effective field theory allows us to systematically focus on the energy range of interest. In the case of Bottomonium \cite{Brambilla:2004jw} we wish to understand the interaction of the medium gluons with the $b\bar{b}$ bound state which occur at the scale of the medium temperature $T\sim 200{\rm MeV}$, much smaller than the rest mass of the individual quarks $m_b=4.66\rm GeV$ \cite{Beringer:1900zz}. Together with the fact the $m_b$ is much larger than the intrinsic scale of QCD $\Lambda_{\rm QCD}\sim 200\rm MeV$ we expect that neither thermal nor quantum fluctuations will be able to pair create heavy $b\bar{b}$ pairs, so that a non-relativistic description is possible \cite{Barchielli:1986zs,Brambilla:1999xf}. 

The starting point for the EFT \cite{Brambilla:1999xf,Rothkopf:2013ria,Ghiglieri:2013iya} is a description of heavy quarkonium in the language of the fundamental field theory QCD, which relies on correlation functions of meson operators $M({\bf x},{\bf y},t)=\Psi({\bf x},t)\Gamma U({\bf x},{\bf y}) \bar{\Psi}({\bf y},t)$. In anticipation of the test charge character of the heavy quarks, a finite separation is introduced by hand ${\bf r}={\bf x}-{\bf y}$, which requires us to connect the fields with a straight Wilson line $U(x,y)$ if gauge invariance shall be preserved in the following. $\Gamma$ denotes a Dirac matrix used to select the appropriate spin structure of the meson, e.g. $\Gamma=\gamma^\mu$ for the vector channel particles.

The quantity we choose to help us understand the in-medium evolution of Bottomonium is the real-time forward propagator 
\begin{align}
 D^>({\bf r},t)=\langle \int D\Psi D\bar{\Psi}\; M({\bf x},{\bf y},t)M^\dagger({\bf x},{\bf y},0) \; e^{iS_{\Psi\bar{\Psi}}[\Psi,\bar{\Psi},A]} \rangle_{\rm medium} \label{Eq:ForwCorr}
\end{align}
It encodes the correlations between a color singlet bare meson inserted into the medium at $t=0$ and its counterpart after having evolved up to time $t$. While initial and final state are devoid of color, intermediate states contributing to Eq.\eqref{Eq:ForwCorr} can have finite color charge, as medium gluons will be absorbed and reemitted over time. If $D^>({\bf r},t)$ is evaluated as follows these medium interactions will be included.

To simplify the description we can exploit the separation of scales by integrating out from the above expression the hard energy scales of the rest masses. To this end we can expand the heavy quark action $S_{\Psi\bar{\Psi}}$ in terms of increasing powers of $m_q^{-1}$. To first order this leads to a decoupling of the upper and lower components of the original Dirac four spinor $\Psi=(\xi,\chi)$ into two two-component Pauli spinors representing the quark and anti-quark respectively. Rewritten in these non-relativistic degrees of freedom, we obtain the heavy quark action of the effective field theory NRQCD. Note that if the heavy mass expansion is continued to higher orders the residual influence of the higher lying scale eventually needs to be taken into account by correctly matching the Wilson coefficients that accompany each term of the NRQCD action \cite{Brambilla:2004jw}.

The quadratic structure of the NRQCD action allows us to integrate out the heavy fermion fields explicitly \cite{Barchielli:1986zs,Rothkopf:2013ria}. Subsequently it becomes possible to leave the language of fields all together and represent the forward correlator in terms of quantum mechanical path integrals, where the fluctuating path ${\bf z}_i(t)$ and the conjugate momenta ${\bf p}_i(t)$ of the quark and antiquark are the dynamical degrees of freedom.  
\begin{align}
 D^>({\bf r},t)=&e^{-2im_Qt} \int_{{\mathbf x}_1}^{{\mathbf x}_2} {\cal D}[{\bf z}_1,{\bf p}_1]\int_{{\mathbf y}_1}^{{\mathbf y}_2} {\cal D}[{\bf z}_2,{\bf p}_2]\; \label{Eq:FullQMPathInt} \\
\nonumber& \times{\rm exp}\Big[ i\sum_{l=1}^2\int_0^{t} ds \Big(  {\bf p}_l(s) \dot{{\bf z}}_l(s) + \frac{{\bf p}_l^2(s)}{2m}\Big)\Big] \left\langle {\cal T} {\rm exp} \Big[  \frac{ig}{c} \oint dx_\mu A_\mu(x)\Big] \right\rangle_{\rm medium},
\end{align}
Note that for static quarks the term on the right is nothing but a Wilson loop in Minkowski time along the rectangular path the heavy quarks trace out in time. In that case the forward propagator obeys a Schroedinger equation
\begin{align}
 i\partial_t D^>({\bf r},t) = \Phi({\bf r},t)D^>({\bf r},t) \label{Eq:StatSchroed}
\end{align}
and we can identify
\begin{align}
  W_\square({\bf r},t)=\left\langle {\cal T} {\rm exp} \Big[  ig \int_\square dx_\mu A_\mu(x)\Big] \right\rangle = {\rm exp}\Big[-i\int_0^t \; ds\; \Phi({\bf r},s)\Big]\label{Eq:WLoopPot}.
\end{align}
We find \cite{Burnier:2012az} that the potential term $\Phi({\bf r},t)$ can actually be a time dependent quantity, which relaxes to a constant only at late times
\begin{align}
 V^{\rm EFT}(r)=\lim_{t\to\infty}\Phi({\bf r},t)=\lim_{t\to\infty}\frac{i\partial_t W_\square({\bf r},t)}{W_\square({\bf r},t)}\label{Eq:DefPot},
\end{align}
since the (instantaneous) effective potential $V^{\rm EFT}({\bf r})$ replaces a retarded interaction mediated by gluons.

Evaluating the real-time Wilson loop and thus $V^{\rm EFT}({\bf r})$ in perturbation theory \cite{Laine:2006ns,Beraudo:2007ky,Brambilla:2008cx}, in AdS/CFT \cite{Hayata:2012rw} and on the lattice \cite{Rothkopf:2009pk,Rothkopf:2011db,Burnier:2013nla} revealed that in general it takes on complex values at finite temperature. In the QGP, i.e. for $T>T_C$ the real-part shows the characteristic Debye screening behavior induced by the presence of deconfined color charges, as one would expect from the analogy with the electromagnetic plasma. On the other hand, since the medium partons are very light they can easily scatter with the gluons that mediate the strong force between the heavy quarks (Landau damping), which is reflected in the presence of an imaginary part that increases with temperature.

This tells us that the QCD derived potential is qualitatively different from purely real potential models, such as the color singlet free or internal energies \cite{Nadkarni:1986as,Satz:2005hx,Kaczmarek:1900zz}, which have been used extensively before the effective field theory approach matured. I.e. even if the real part of the potential is quantitatively captured in a satisfying manner by the model potential \cite{Burnier:2009bk,Burnier:2013nla}, the absence of the imaginary part is a sign of missing physics.

The fact that the inter-quark potential defined from an effective field theory description is complex urges us to carefully examine its meaning. The central observation to make is that the Schroedinger equation \eqref{Eq:StatSchroed} does not act on the wave-function $\psi_{Q\bar{Q}}({\bf r},t)$ of the two-body system but instead operates on the in-medium correlation function $D^>({\bf r},t)$.\footnote{This difference is also reflected in the initial conditions appropriate for \eqref{Eq:StatSchroed}. We have $D^>({\bf r},0)=\delta({\bf r})$ independent of the initial wavefunction of the system.} Hence the presence of an imaginary part in $V_{\rm EFT}({\bf r})$ does not make any immediate statement on a reduced probability to find the heavy Quarkonium or in other words on the decay of the $Q\bar{Q}$ pair. In particular the infinitely heavy quark and anti-quark pair for which the complex potential has been originally defined, will always remain present in the system, as its decay into gluons is fully 
suppressed in the 
non-relativistic description. 

${\rm Im}[V^{\rm EFT}({\bf r})]\neq0$ instead reflects the fact that the thermal fluctuations of the medium decorrelate the current state from its initial conditions. Note that this too is not a statement on the melting of heavy quarkonium states, but only on the disappearance of correlations. To adequately take into account this phenomenon of thermal decoherence in the description of heavy quarkonium evolution, we turn to an approach based on an open-quantum systems viewpoint.

\subsection{An Open-Quantum systems viewpoint}

The idea of treating heavy quarkonium as an open-quantum system has received increasing attention in recent years \cite{Young:2010jq,Borghini:2011yq,Borghini:2011ms,Akamatsu:2011se,Dutta:2012nw,Akamatsu:2012vt}. The formalism originally developed in the context of condensed matter systems \cite{Breuer:2002pc} is particularly suited to the study of Bottomonium since the intrinsic separation of scales allows a clear distinction between environment (QGP) and test system (Bottomonium) degrees of freedom. Our aim is to use the stochastic potential approach to heavy quarkonium evolution, first introduced in \cite{Akamatsu:2011se}.  It makes close contact to the potentials derived in the EFT's and by doing so allows us to incorporate non-perturbatively information about the medium if e.g. potentials extracted from lattice QCD (see e.g. \cite{Rothkopf:2011db,Burnier:2013nla}) are used.

The overall system consisting of the environment, represented by the QGP, and the $Q\bar{Q}$ test system is described by a hermitian Hamiltonian $H_{\rm full}=H^\dagger_{\rm full}$
\begin{align}
 H_{\rm full}=H_{\rm med}\otimes I_{Q\bar{Q}}+ I_{\rm med}\otimes H_{Q\bar{Q}}+H_{\rm int}, \quad \frac{d}{dt}\sigma(t)=-i[H_{\rm full},\sigma(t)],
\end{align}
The states of the system evolve unitarily under $H_{\rm full}$ and the density matrix of states $\sigma(t)$ follows the von-Neumann equation. For simplicity we assume a separation of the initial states $\sigma(0)=\sigma_{\rm med}\otimes\sigma_{Q\bar{Q}}(0)$. In the following we wish to track the evolution of the $Q\bar{Q}$ subsystem by using only its own degrees of freedom , which is why we proceed by tracing out all the medium degrees of freedom. One arrives at the following heavy quarkonium density matrix
\begin{align}
 \sigma_{Q\bar{Q}}(t,{\bf r}, {\bf r}')={\rm Tr}_{\rm med}\Big[\sigma(t,{\bf r},{\bf r}')\Big]=\langle \Psi_{Q\bar{Q}}({\bf r},t)\Psi_{Q\bar{Q}}^*({\bf r}',t)\rangle \label{Eq:QQDensMat}.
\end{align}
where $\psi_{Q\bar{Q}}({\bf r},t)$ denotes the wavefunction of heavy quarkonium two-body system in relative coordinates.

The phenomenon of decoherence formally tells us that \cite{Breuer:2002pc} there exists an apriori unknown basis of states in which the reduced density matrix of states Eq.\eqref{Eq:QQDensMat} becomes diagonal over time. The selection of this basis as well as the timescale for decorrelation depends on the interaction between the medium and the test-system. Intuitively decoherence is related to the fact that thermal fluctuations distort the wavefunction of the test-system as it evolves in time, so that taking the thermal average over many realizations of the system leads to a loss of norm in the averaged wave-function.

We follow \cite{Akamatsu:2011se} in setting up an description of the time evolution of the static quarkonium system on the level of its wavefunction $\psi_{Q\bar{Q}}({\bf r},t)$. Our goal is to both implement the concept of decoherence encoded in the imaginary part of $V^{\rm EFT}({\bf r})$ while still preserving the purely real values of the energy for the two-body system. 
In particular this prohibits us from directly using ${\rm Im}[V^{\rm EFT}]({\bf r})$ in the Hamiltonian of the system.\footnote{This difference in the interpretation of ${\rm Im}V^{\rm EFT}({\bf r})$ distinguishes our study from e.g. \cite{Margotta:2011ta,Strickland:2011mw,CasalderreySolana:2012av}, where a complex potential has been used explicitly to evolve the wave function. This non-hermiticity leads e.g. to the presence of an imaginary part in the binding energy, contrary to the open-quantum systems approach laid out in this section, where all energies of the non-relativistic heavy quarkonium system remain real.} A rigorous 
derivation of the stochastic potential, as well as its generalization to finite mass using the Feynman-Vernon influence functional has been worked out in ref.\cite{Akamatsu:2012vt}, .

The idea behind our approach is to reinterpret the presence of the imaginary part of the EFT potential $V^{\rm EFT}({\bf r})$ as an uncertainty in the values of a purely real potential $V_{Q\bar{Q}}({\bf r})$ governing the evolution of $\psi_{Q\bar{Q}}({\bf r},t)$. I.e. once a bound state of heavy quarks held together by a color string enters the QGP it will experience a weakening of the force between its constituents (Debye screening) and the gluons of the medium will continuously kick the color string (Landau damping) leading to slightly different values of the screened potential at subsequent times.

A naive way to formalize this idea is to construct a fully hermitian time evolution operator
\begin{align}
 \psi_{Q\bar{Q}}({\mathbf r},t)={\cal T} {\rm exp}\Bigg[ -i\int_0^t\;ds\;\Big\{ -\frac{\nabla^2}{m_Q}+2m_Q+ V_{Q\bar{Q}}({\mathbf r})+\Theta({\mathbf r},s)\Big\}\Bigg]\psi_{Q\bar{Q}}({\mathbf r},0)\label{Eq:StochTimeEvol},
\end{align}
whose real potential term is disturbed by Gaussian, i.e. Markovian noise $\langle\Theta({\mathbf r},t)\rangle=0$. In order to encode the thermal properties of the QGP medium we allow the noise to carry a non-trivial spatial correlation structure $\langle \Theta({\mathbf r},t)\Theta({\mathbf r',t'})\rangle=\frac{1}{\Delta t}\delta_{t,t'} \Gamma({\mathbf r},{\mathbf r}')$.

Expanding Eq.\eqref{Eq:StochTimeEvol} according to the rules of Ito calculus, we obtain the following stochastic Schroedinger-type equation
\begin{align}
\hspace{-0.1cm} i\frac{d}{dt}\psi_{Q\bar{Q}}({\mathbf r},t)=\Big(-\frac{\nabla^2}{m_Q}+2m_Q+V_{Q\bar{Q}}({\mathbf r})+\Theta({\mathbf r},t)-i\frac{\Delta t}{2}\Theta^2({\mathbf r},t)\Big)\psi_{Q\bar{Q}}({\mathbf r},t)\label{Eq:StochSchroed}.
\end{align}
Note that the term on the RHS with a factor $i$ originates from the fluctuations $\Theta({\bf r},t)$. Evolving a single realization of the heavy quarkonium system, i.e. an individual wavefunction $\psi_{Q\bar{Q}}({\bf r},t)$, along real-time still preserves its norm $|\psi_{Q\bar{Q}}|(t)=1$. Contrast this to the behavior after taking the thermal average
\begin{align}
 i\frac{d}{dt}\langle\psi_{Q\bar{Q}}({\mathbf r},t)\rangle=\Big(-\frac{\nabla^2}{m_Q}+2m_Q+V_{Q\bar{Q}}({\mathbf r})-\frac{i}{2} \Gamma({\mathbf r},{\mathbf r})\Big)\langle\psi_{Q\bar{Q}}({\mathbf r},t)\rangle\label{Eq:StochAvgSchroedinger}.
\end{align}
It turns out that the averaged wave function $\langle\Psi_{Q\bar{Q}}({\mathbf r},t)\rangle$ or equivalently the thermally averaged correlator $\langle\Psi_{Q\bar{Q}}({\mathbf r},t)\Psi^*_{Q\bar{Q}}({\mathbf r},0)\rangle$ obeys a Schroedinger equation with an imaginary part. In addition Eq.\eqref{Eq:StochAvgSchroedinger} intimately links the appearance of a non-hermiticity to the diagonal structure of the noise correlations. If we identify this $\Gamma({\bf r},{\bf r})$ with the imaginary part of $V_{\rm EFT}({\bf r})$ we have established a direct link between the EFT language and the open-quantum systems picture, where the thermal fluctuations of the medium are characterized via noise correlations.

Note that from this viewpoint the information obtained from $V_{\rm EFT}({\bf r})$ is not yet enough to fully specify the dynamics of the system, indeed up to now, the off-diagonal parts of $\Gamma({\bf r},{\bf r}')$ are not determined. Since Eq.\eqref{Eq:StochAvgSchroedinger} does not contain further reference to the off-diagonal terms we might ask, whether such additional information is actually relevant to understanding Bottomonium evolution? 

\begin{figure}
\centering
\includegraphics[scale=0.6, clip=true, trim=5cm 8cm 5cm 6cm]{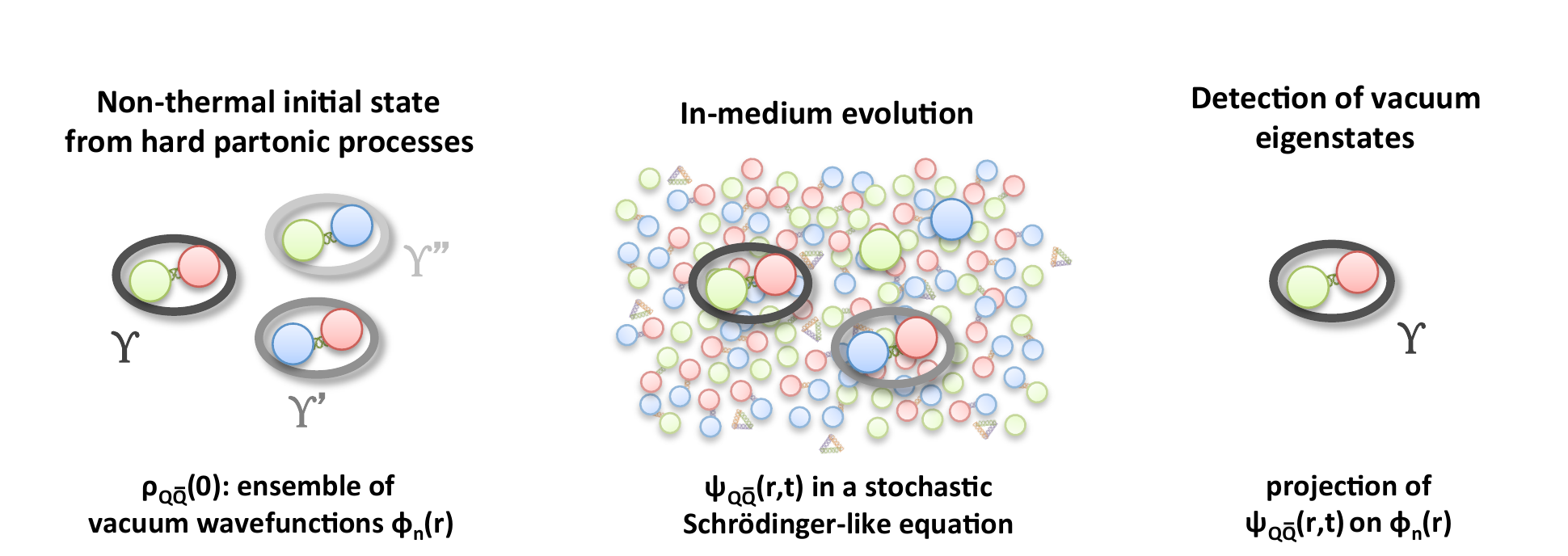}
\caption{A schematic view of the setup used to describe Bottomonium evolution.}\label{Fig:Schematic}
\end{figure}

Up to now we have only spoken about the wavefunction of the heavy quarkonium system. To make a statement relevant for phenomenology, we however need to know the abundances of the different Bottomonium states present in the medium after a certain time t. In particular we ask (Fig.~\ref{Fig:Schematic}): what is the probability to find an eigenstate $\phi_n({\bf r})$ of the vacuum Hamiltonian $$H^{\rm VAC}_{Q\bar{Q}}\phi_n({\bf r})=E_n\phi_n({\bf r})$$ at time $t$, as it is the remnants of such states which are measured in experiment. In the stochastic potential approach the answer lies in the so called state admixture or survival probability of the state $\phi_n({\bf r})$ denoted by $c_{nn}(t)$. It is obtained by projecting the current wavefunction $\psi_{Q\bar{Q}}({\bf r},t)$ onto the corresponding vacuum wavefunction $\phi_n({\bf r})$ 
\begin{align}
 c_{nn}(t)=\int d^3r\,d^3r' \; \phi^*_n({\mathbf r})\langle \Psi_{Q\bar{Q}}({\mathbf r},t) \Psi^*_{Q\bar{Q}}({\mathbf r}',t)\rangle \phi_n({\mathbf r}').\label{Eq:DefAdmix}
\end{align}
Note that this quantity depends on the evolution of the density matrix \linebreak $\sigma_{Q\bar{Q}}(t,{\bf r},{\bf r}')=\langle \Psi_{Q\bar{Q}}({\mathbf r},t) \Psi^*_{Q\bar{Q}}({\mathbf r}',t)\rangle$ whose time dependence is actually sensitive to the off-diagonal elements of $\Gamma({\bf r},{\bf r}')$ (see Sec.~IV in \cite{Akamatsu:2011se}). This tells us that observing the evolution of Bottomonium gives us access to information that goes beyond what we can obtain from the EFT potential and thus by extension the thermal spectral functions.

The difference to inserting an imaginary part directly into the Schroedinger equation of $\psi_{Q\bar{Q}}({\bf r},t)$ is now also evident. Instead of calculating the time evolution of the fluctuating $\psi_{Q\bar{Q}}({\bf r},t)$ one actually determines the time evolution of $\langle \psi_{Q\bar{Q}} \rangle({\bf r},t)$ which however does not allow us to extract the time evolution of states in Eq.\eqref{Eq:DefAdmix}, since in general $$\langle \psi_{Q\bar{Q}} \rangle\langle \psi_{Q\bar{Q}} \rangle^*\neq \langle \psi_{Q\bar{Q}} \psi_{Q\bar{Q}}^* \rangle$$

We have to caution the reader however that the intuitive setup presented here of course has its own deficiencies. The possibility to describe the heavy quarkonium evolution solely by a real stochastic potential has been shown to be valid only in the infinite mass case. By allowing a finite mass term in Eq.\eqref{Eq:StochTimeEvol} we venture beyond its formal applicability, since the quantum analog of the classical drag force is not taken into account. The system described by Eq.\eqref{Eq:StochSchroed} thus cannot reach thermal equilibrium even at asymptotically late times and will eventually accumulate energy at a linear rate in time\cite{Akamatsu:2011se,Akamatsu:2012vt}. Since our goal here lies in understanding relatively early time physics $t\leq 2 \rm fm$ where the artificial linear energy rise is not yet dominating, we believe that the numerical simulations presented in the following section do contain insightful information on the Bottomonium evolution.

\section{Bottomonium melting via a stochastic potential}

By constructing an approximate open-quantum systems description of heavy quarkonium based on the concept of stochastic potential in the previous section we are able to relate the real- and imaginary part of the EFT potential $V^{\rm EFT}({\bf r})$ to the heavy quarkonium potential $V_{Q\bar{Q}}({\bf r})$ and the diagonal noise correlations of the medium $\Gamma({\bf r},{\bf r})$. In the following we will perform numerical simulations based on Eq.\eqref{Eq:StochSchroed} and investigate the real-time evolution of Bottomonium at high temperatures to obtain a first view on the melting process from the stochastic potential approach.

We describe the vacuum properties of the Bottomonium system by utilizing a Cornell type potential 
\begin{align}
 V^{\rm VAC}({\bf r})=\sigma |{\bf r}| - \frac{\alpha}{|{\bf r}|}, \qquad \sigma=(0.42)^2 {\rm GeV},\; \alpha=0.1\label{Eq:VacPot}
\end{align}
in the Hamiltonian 
\begin{align}
 H^{\rm VAC}=2m_b-\frac{\nabla^2}{m_b}+V^{\rm VAC}({\bf r}), \quad m_b=4.5663{\rm GeV}
\end{align}
with the bare bottom mass $m_b$, the string tension $\sigma$ and the strong coupling $\alpha$ chosen such that the ground state Upsilon has a mass of $m_\Upsilon=9.482\rm GeV$. Letting the linear term in Eq.\eqref{Eq:VacPot} go over smoothly to a constant value at $r_{SB}=1.32\rm fm$ to take into account QCD string breaking leads to a spectrum in which there exist three bound S-states $(\Upsilon,\Upsilon',\Upsilon")$ and e.g. three bound P-states $(^1\chi_b,^2\chi_b,^3\chi_b)$ and two bound D-states $(^1\Upsilon^3D_2,^2\Upsilon^3D_2)$ below the B-meson threshold.

\begin{figure}
 \includegraphics[angle=-90, scale=0.3]{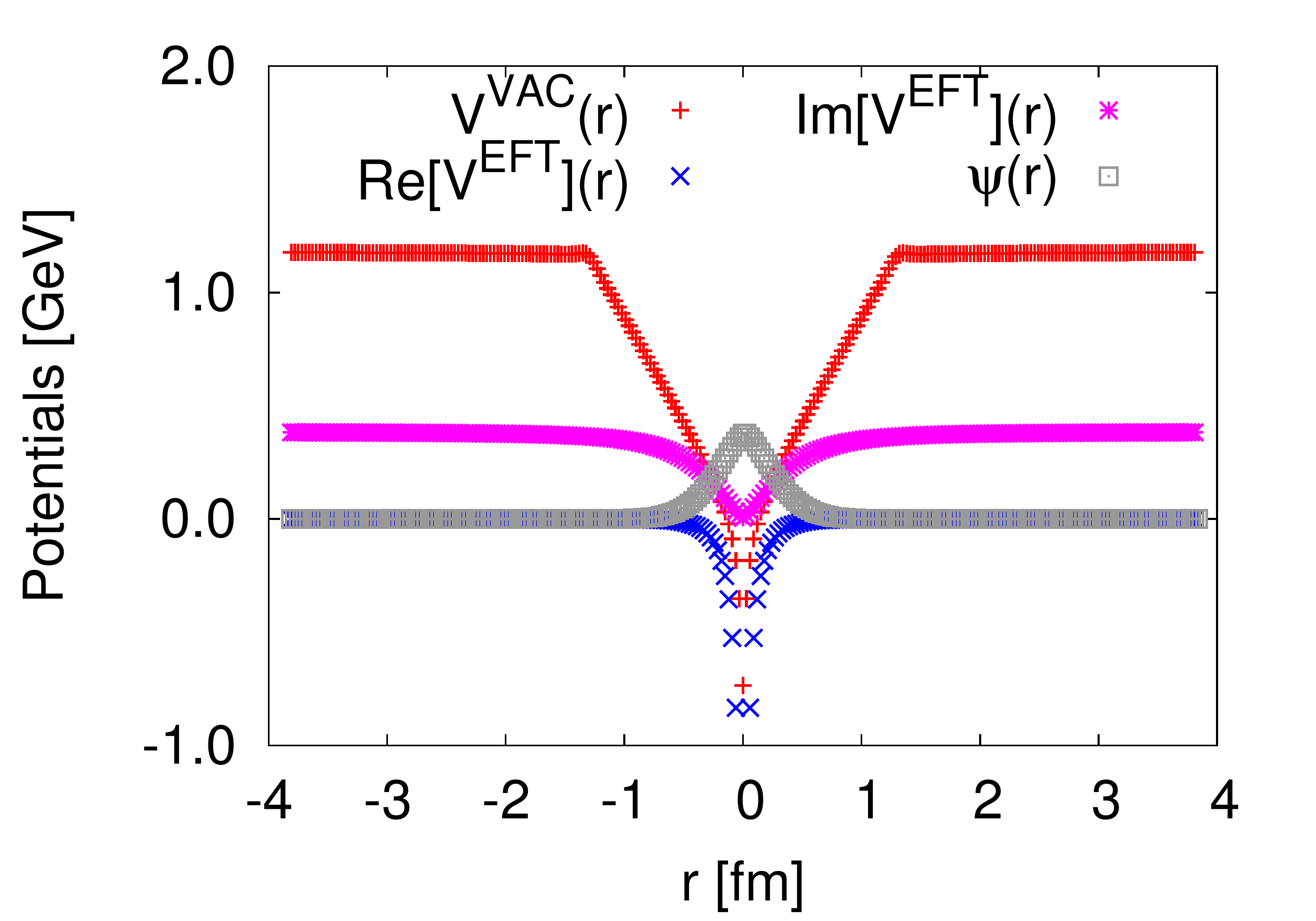}
 \includegraphics[angle=-90, scale=0.3]{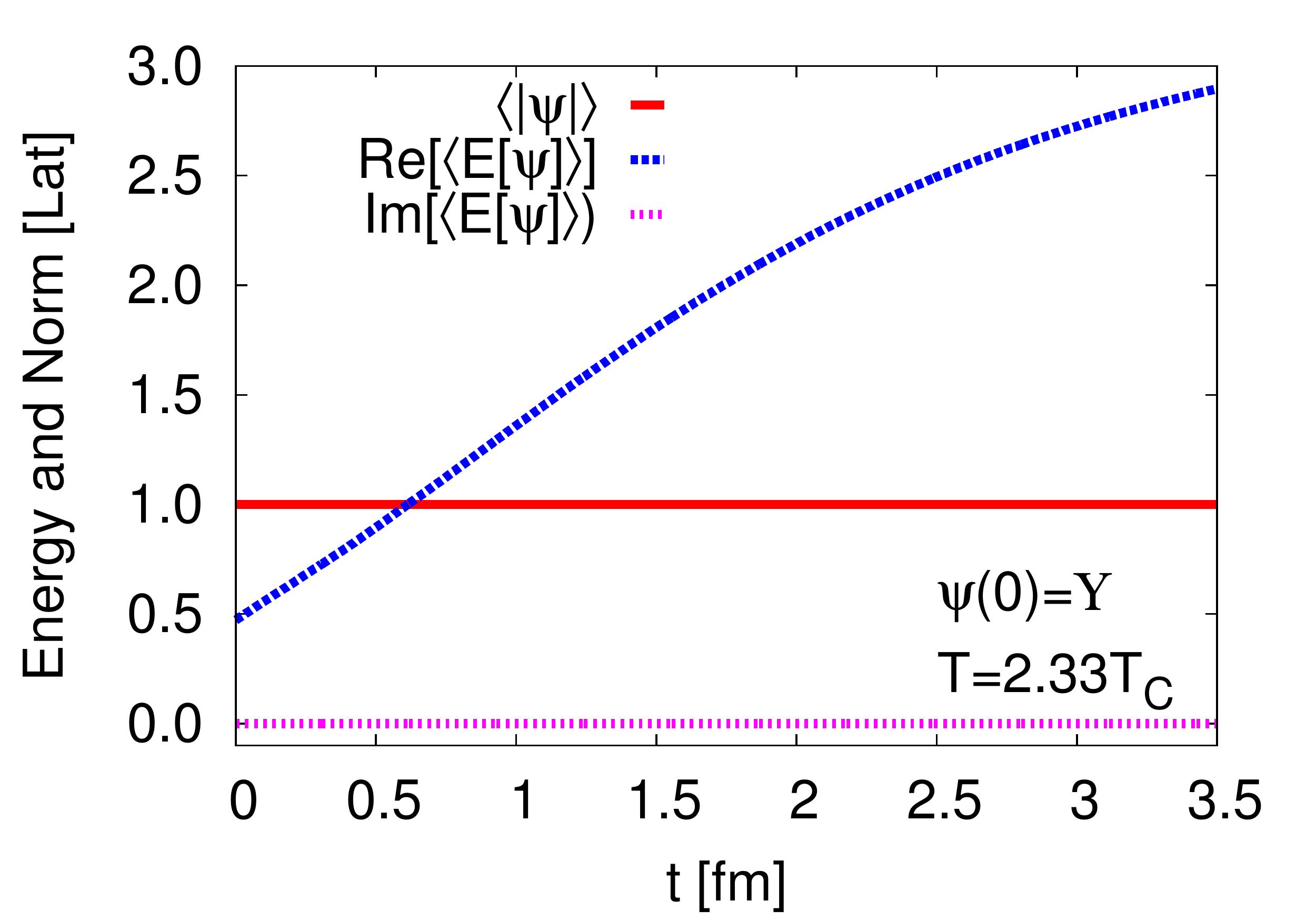}
 \caption{(left) The radial component of the vacuum potential (red) used to define the vacuum Bottomonium states. The (blue) curve denotes the real part of the in-medium heavy quark potential $V_{Q\bar{Q}}({\bf r})$ from the real part of the HTL potential at $T=2.33T_C=396\rm MeV$, while we show the values of the imaginary part of the EFT potential in (magenta). (right) Check of the unitary nature of the time evolution in Eq.\eqref{Eq:StochSchroed} as the wavefunction norm (red) remains at unit value and the energy of the system remains purely real. (blue) $\langle E[\psi]\rangle$ rises due to the influx of energy from the thermal medium but due to the absence of the correct momentum drag it will not settle to a time independent value and an artificial linear rise will dominate at later times \cite{Akamatsu:2012vt}.}
 \label{Fig:PotentialsNrg}
\end{figure}
\begin{figure}
\centering
 \includegraphics[angle=-90, scale=0.3]{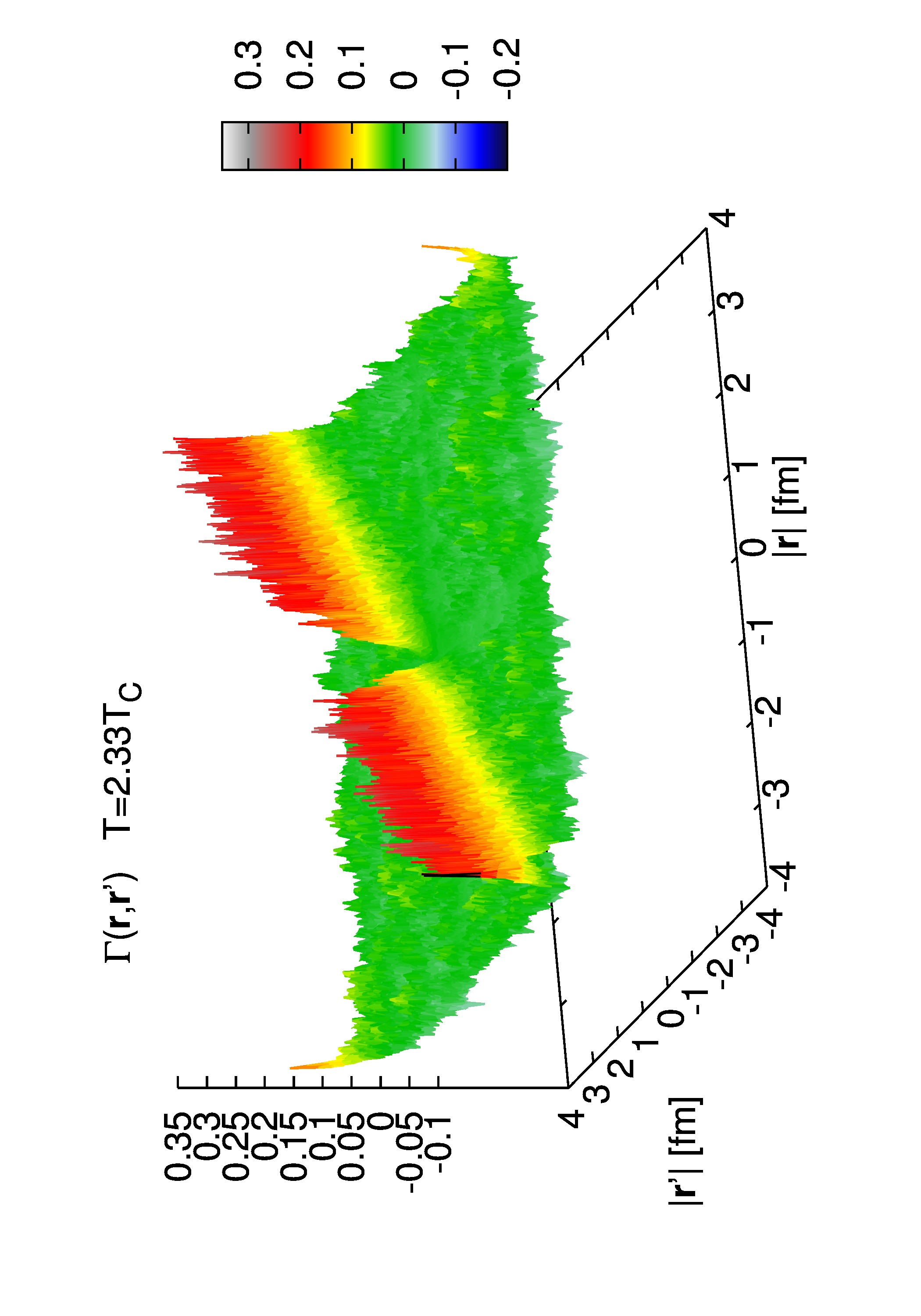}
 \caption{ The actual correlation function $\Gamma({\bf r},{\bf r}')$ obtained from sampling 60 noise realizations. Note that the diagonal follows the values of the imaginary part of the EFT potential $\Gamma({\bf r},{\bf r})=-{\rm Im}[V^{\rm EFT}]({\bf r})$ and the transversal falloff obeys a Gaussian  $\Gamma({\bf r},{\bf r}')\propto {\rm exp}[-|{\bf r}-{\bf r}'|^2/2\lambda^2]$ with characteristic length $\lambda=1/T$.}
 \label{Fig:NoiseCorrRealiz}
\end{figure}

For the description of the in-medium evolution we have to specify the real values of the heavy-quark potential $V_{Q\bar{Q}}({\bf r})$ as well as the correlations of the thermal noise $\Gamma({\bf r},{\bf r}')$. Since in this exploratory study we will work at rather high temperatures $T=2.33T_C=396\rm MeV$ we rely on the perturbative values of the EFT potential shown in Fig.~\ref{Fig:PotentialsNrg}. The functional form taken from \cite{Laine:2006ns} reads
\begin{align}
 V_{Q\bar{Q}}({\bf r})&=-\frac{3g^2 }{4\pi}\frac{e^{-m_D |{\bf r}|}}{|{\bf r}|}\\
 \Gamma({\bf r},{\bf r})&= - \frac{3g^2T}{4\pi} \phi(m_D |{\bf r}|),\quad \phi(x)=2\int_0^\infty dz \frac{z}{(z^2+1)^2}\Big[1-\frac{sin[zx]}{zx}\Big], \label{Eq:LainePot}
\end{align}
where we set the Debye mass to $m_D=1\rm GeV$ and the gauge coupling to $g=2.14$.

To specify the off-diagonal elements of the noise we make the simple Ansatz that they decay in a Gaussian fashion with a characteristic correlation length given by $\lambda=1/T$ (see e.g. Fig.~\ref{Fig:NoiseCorrRealiz})
\begin{align}
 \Gamma({\bf r},{\bf r}')=\sqrt{\Gamma({\bf r},{\bf r})\Gamma({\bf r}',{\bf r}')}{\rm exp}[-|{\bf r}-{\bf r}'|^2/2\lambda^2]
\end{align}
While this choice seems reasonable at high temperatures, the behavior of the phenomenologically relevant QGP close to the phase transition (or rather cross-over) is characterized by growing correlation lengths and deviations from a simple Gaussian must be expected. Ultimately we will need to to find ways both conceptually and technically how $\Gamma({\bf r},{\bf r}')$ can be determined from an appropriate lattice QCD observable in its entirety.

With these prerequisites set, we proceed by discretizing the system on a three dimensional hypercube with $N=256$ points along each axis. The physical length of the box is chosen to be $L_{\rm box}=7.68\rm fm$ with a lattice spacing of $\Delta x=0.03\rm fm$ so that all three $b\bar{b}$ S-wave bound state wavefunctions both fit into the available volume and are sufficiently well resolved (see gray curve on the left in Fig.~\ref{Fig:PotentialsNrg}). To numerically obtain these eigenfunctions for the vacuum potential of Eq.\eqref{Eq:VacPot}, which we will use as initial conditions for the in-medium evolution, we rely on the PETSC \cite{PETSC} and SLEPC \cite{SLEPC} library. 

The dynamics of the Bottomonium states according to Eq.\eqref{Eq:StochSchroed} is subsequently determined using the unconditionally stable and norm preserving Crank-Nicolson method \cite{NR} with a time step of $\Delta t=7.5\times 10^{-4}\rm fm$. We prepare ten realizations of the system based on different initial conditions for the noise generator. In each run the norm of the wavefunction remains at unity as shown in the right panel of Fig.~\ref{Fig:PotentialsNrg}. For every fourth step in time we determine the admixtures of the individual bound states $c_{nn}(t)$ by projecting the stochastically evolved wavefunction onto the Eigenfunctions $\phi_n({\bf r})$ according to Eq.\eqref{Eq:DefAdmix} and carrying out the noise average. 

\begin{figure}[t]
 \includegraphics[angle=-90, scale=0.15]{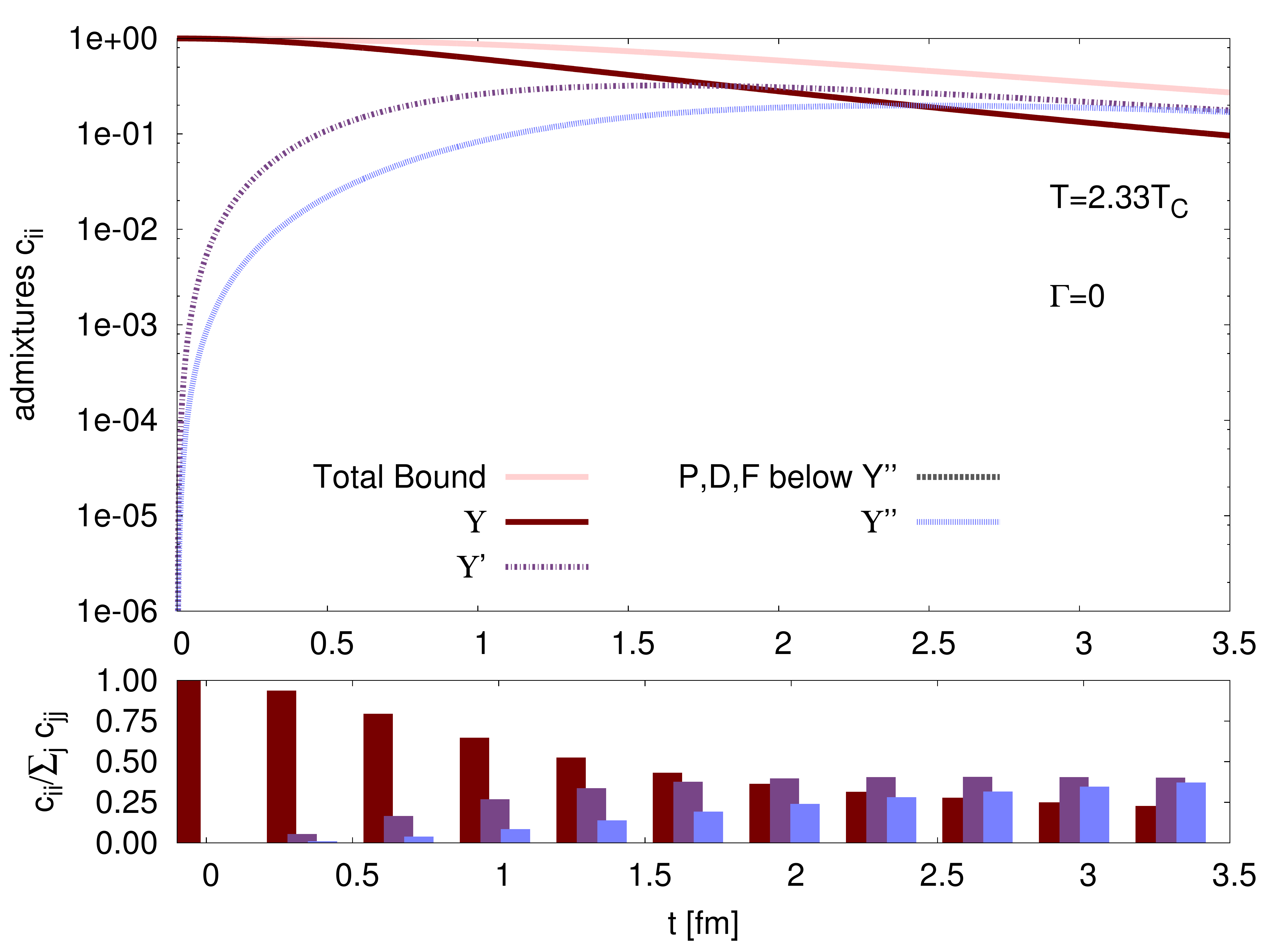}
 \includegraphics[angle=-90, scale=0.15]{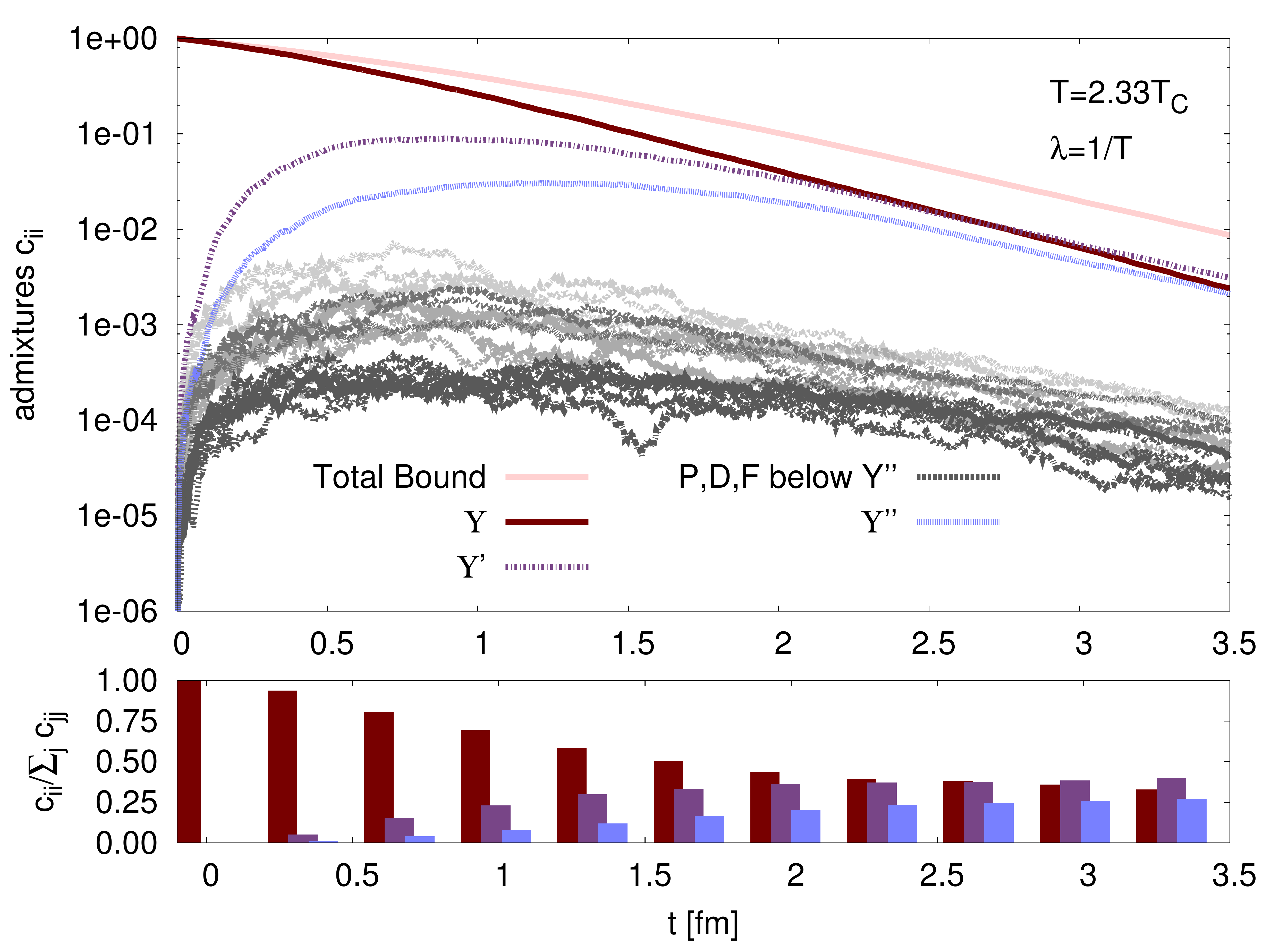}
 \caption{Time evolution of the absolute (top) and relative (bottom) abundances $c_{ii}$ of different Bottomonium bound states in the QGP at $T=2.33T_C$ starting from a pure Upsilon $\psi_{Q\bar{Q}}({\bf r},0)=\phi_0({\bf r})$ configuration. The values of the real inter-quark potential are taken from the perturbative HTL potential $V_{Q\bar{Q}}({\bf r})={\rm Re}[V^{\rm HTL}]({\bf r})$ by Laine \cite{Laine:2006ns}. Distinct pattern emerge between the case where (left) no noise is included in our system $\Gamma({\bf r},{\bf r}')=0$ and (right) the diagonal correlations of the thermal noise resemble the imaginary part of the HTL potential $\Gamma({\bf r},{\bf r})=-{\rm Im}[V^{\rm HTL}]({\bf r})$. In both cases state mixing occurs since the vacuum initial state $\phi_0({\bf r})$ is not an Eigenstate of the in-medium Hamiltonian. Note that state mixing only changes the abundances of states with the same parity, i.e. no P,D or F states appears (left). Thermal noise contributes additional excitations and 
deexcitations to 
the time evolution which (albeit weakly) also excited states with different angular momentum. 
}\label{Fig:BottomStochAndNoNoise}
\end{figure}

\subsection{Pure initial state evolution}

Before investigating the fully stochastic evolution of Bottomonium in a static QGP medium we first wish to observe its behavior in the absence of thermal noise. This case, which is known very well from the study of purely real model potentials (see also \cite{Borghini:2011yq,Borghini:2011ms,Akamatsu:2011se,CasalderreySolana:2012av,Dutta:2012nw}), already allows us to observe an important contribution to the dynamics: state mixing. 
The vacuum $\Upsilon$ initial state is not an Eigenstate of the Hamiltonian with Debye screened potential. Hence as shown on the left of Fig.~\ref{Fig:BottomStochAndNoNoise}, its admixture decreases steadily and instead the excited states $\Upsilon'$ and $\Upsilon"$ become populated. At $t\sim2 \rm fm$ almost equal amounts can be found. Note that there is a distinct time-lag between the increase in the numbers of $\Upsilon'$ and $\Upsilon''$ reflecting their different masses. Summing up the admixtures of the bound states we find that this value deviates slightly from one, the continuum states are only weakly populated. As the Hamiltonian respects parity, we only find mixing between states that have the same angular momentum $L$, i.e. none of the  possible P, D, F or higher $L$ states appear during the evolution.

In the case where we take the effects of thermal decoherence, i.e. of the imaginary part of the EFT potential into account, we still find that state mixing is an important part of the dynamics. As shown on the right of Fig.~\ref{Fig:BottomStochAndNoNoise} the stochastic noise however leads to a much faster depopulation of the $\Upsilon$ ground state. Since the $\Theta({\bf r},t)$ in our approach does not discriminate between states of different angular momentum, it thermally excites all of the different P, D and F states albeit with a small magnitude. The finite angular momentum states of mass below $m_{\Upsilon''}$ with their respective multiplicities $3(P)$, $5(D)$ and $7(F)$ are depicted in the figures as individual gray lines with a common shading\footnote{The roughness but not the overall magnitude of the gray curves is owed to the relatively small number of 10 individual runs used to generate the noise average due to the involved numerical cost.}. Since we do not yet have a baseline for the overall 
production of $b\bar{b}$ in heavy-ion collisions we concentrate here only on the relative abundances between the three S-states plotted in the lower part of the figure. 

\begin{figure}
 \includegraphics[angle=-90, scale=0.15]{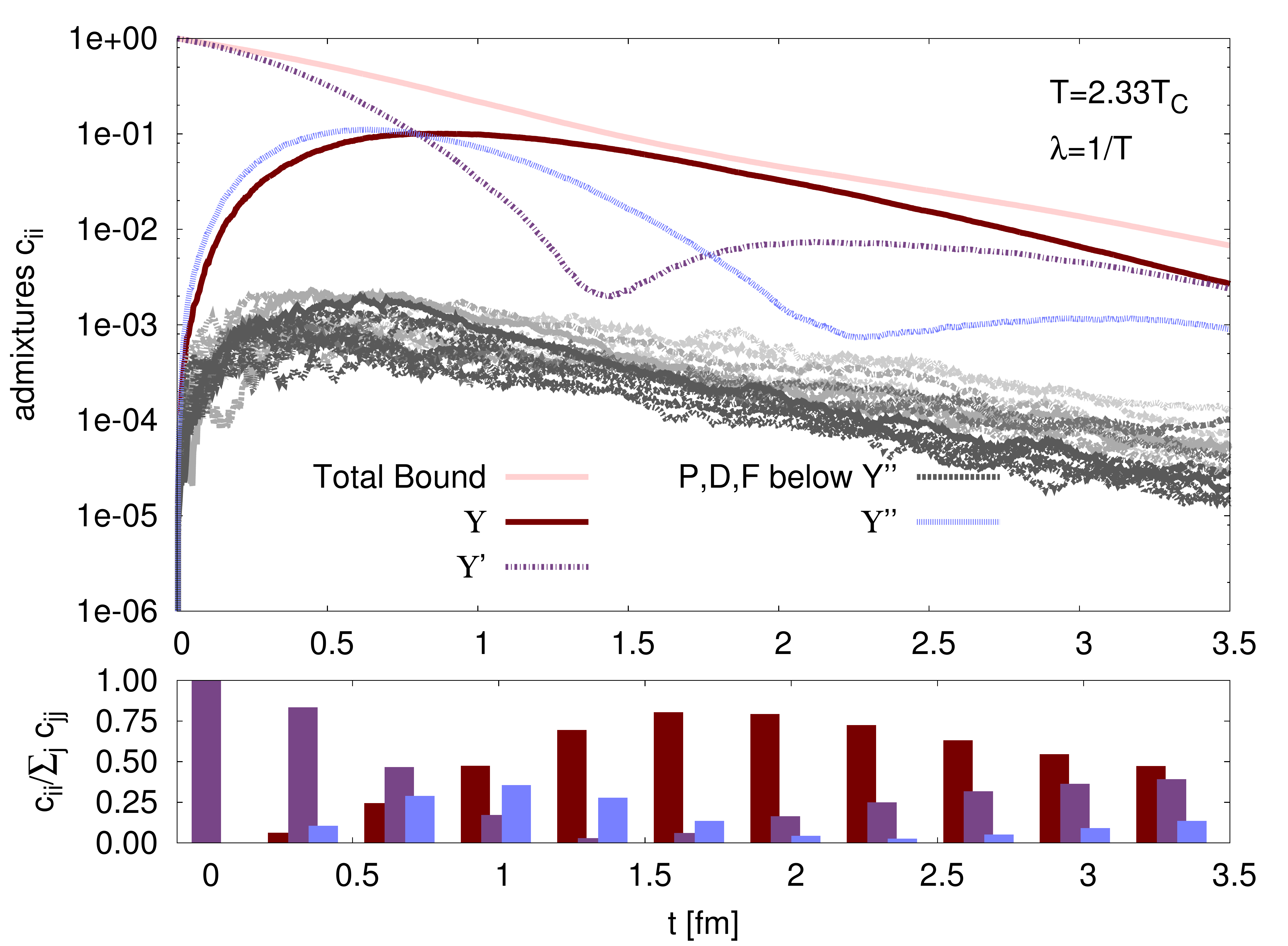}
 \includegraphics[angle=-90, scale=0.15]{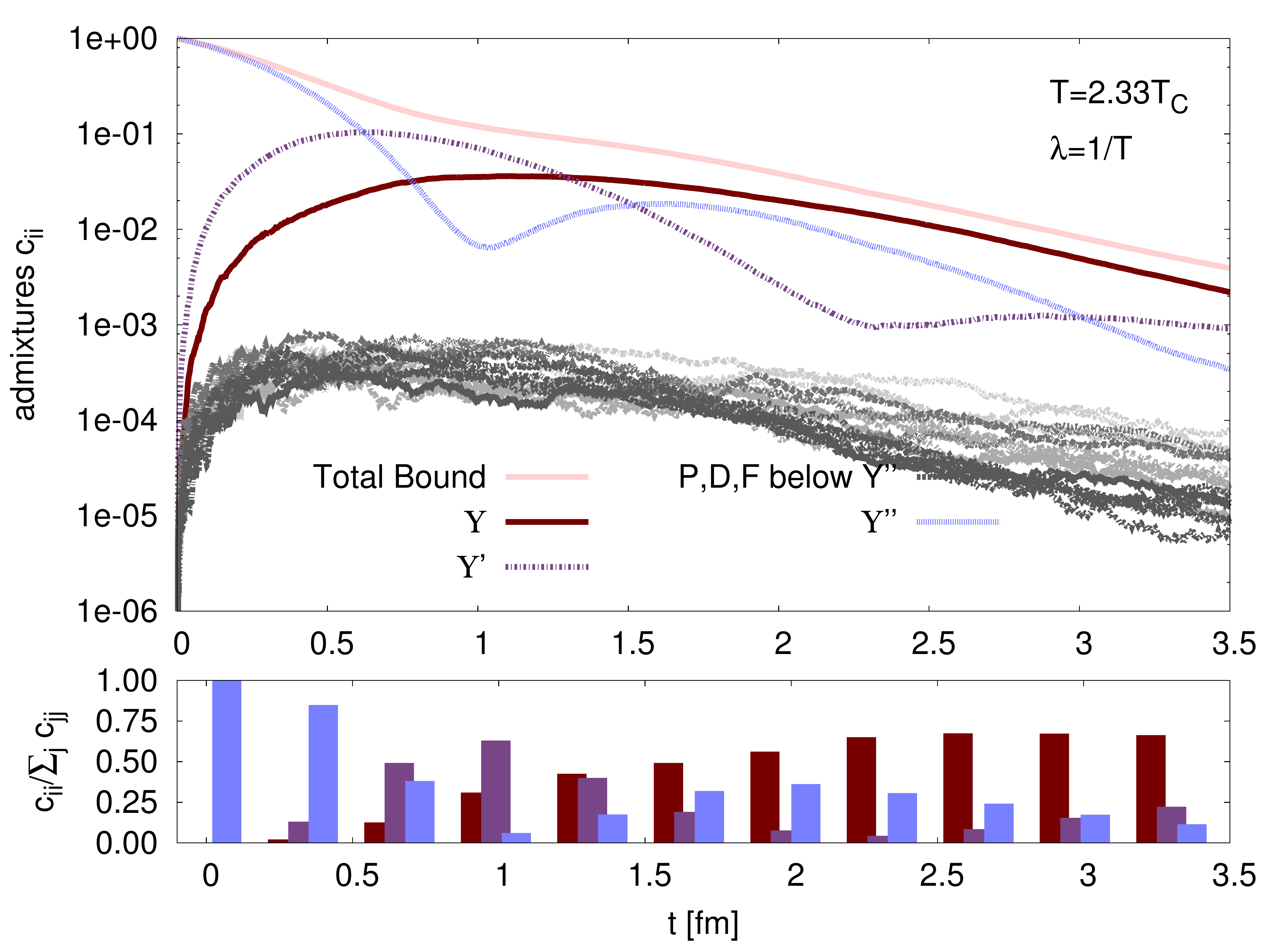}
 \caption{Time evolution of the absolute (top) and relative (bottom) abundances $c_{ii}$ of different Bottomonium bound states in the QGP at $T=2.33T_C$ based on the HTL potential $V_{Q\bar{Q}}={\rm Re}[V^{\rm HTL}]({\bf r})$ and $\Gamma({\bf r},{\bf r})=-{\rm Im}[V^{\rm HTL}]({\bf r})$. Initial conditions are given by a pure (left) $\Upsilon'$ or (right) $\Upsilon ''$ configuration. While $\Upsilon^{\prime\prime}$ appears to be melting fastest compared to the other S-wave Bottomonium states, both excited states contribute to a significant replenishment of the ground state over time. }
 \label{Fig:ExcitedBottom}
\end{figure}

The evolution of initial excited states plotted in Fig.~\ref{Fig:ExcitedBottom} shows much shorter timescales for the initial depopulation as in the case of the ground state. As expected, if we start with $\Upsilon'$ it affects both the bound states above and below, so that the admixtures of  $\Upsilon$ and $\Upsilon "$ up to $t<0.8 \rm fm$ grow with similar speed. Since $\Upsilon"$ however cannot be efficiently fed by the ever declining $\Upsilon'$ it subsequently depopulates again for $t>1\rm fm$. In the late time region $t>1.5\rm fm$ where the quantitative reliability of our simulation is not guaranteed anymore, we find that the buildup in $\Upsilon$ will lead to a replenishment of the almost fully depopulated $\Upsilon'$ state.

As last example in this subsection we take the evolution of an initial $\Upsilon''$ state which exhibits the most rapid depopulation of all three states. The total admixture of bound states exhibits two regimes of distinct rates of decrease in this case. The first is dominated by the faster than exponential decrease of the $\Upsilon ''$, the second, slower one, follows the decay of the ground state which has taken over after $t>1.3\rm fm$.

\subsection{Mixed initial state evolution}

Since in the initial hard partonic stages of the collision at the LHC several $b\bar{b}$ pairs will be produced, we have to take into account the possibility that an ensemble of different Bottomonium states will enter the QGP together. Thus we have now two ensembles to take care of in the simulation, one concerns the realizations of the thermal noise, the other the probability to find a certain Bottomonium state as initial condition.

\begin{figure}
\centering
 \includegraphics[angle=-90, scale=0.15]{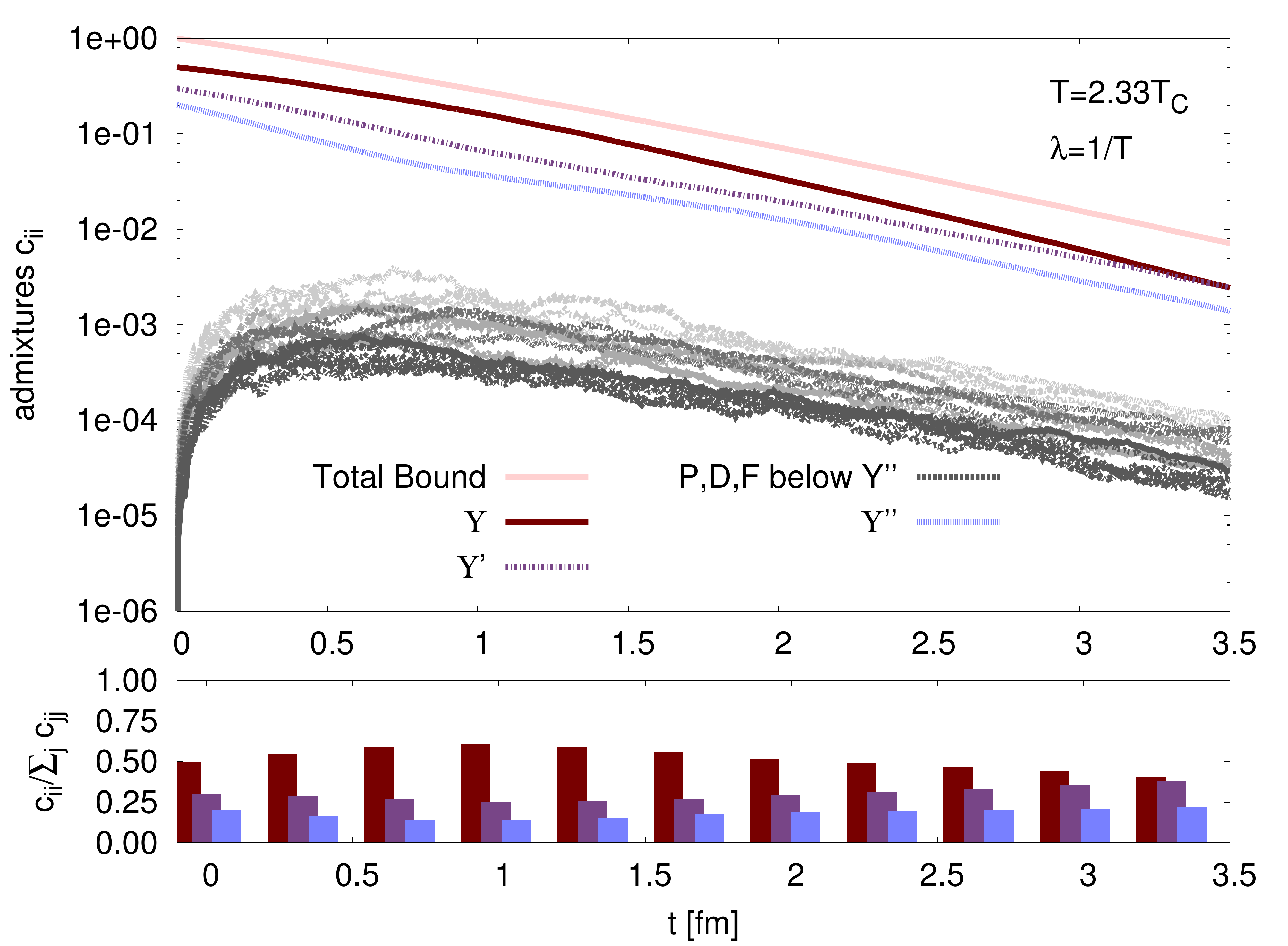}
 \caption{The evolution pattern for the Bottomonium family in the presence of a mixed initial state. Here we use relative abundances inferred from the CMS dilepton spectrum in $\rm p+p$, i.e. $\Upsilon:\Upsilon':\Upsilon" = 5:3:2$ as starting point for the evolution in a medium at $T=2.33T_C$. In this highly idealized setting, we observe that the interplay between replenishment and suppression of the individual Bottomonium states for times $t<1.5\rm fm$ actually leads to a slight enhancement of the ground state compared to the excited states.  }
 \label{Fig:CMSinitCondEvol}
\end{figure}

Here we make a naive ansatz and roughly take the relative abundances of the Bottomonium S-wave states obtained from the measurements carried out by the CMS collaboration \cite{Chatrchyan:2011pe} for $\rm p+p$ collisions at $\sqrt{s}=2.67\rm TeV$. Using the ratio $\Upsilon:\Upsilon':\Upsilon'' = 5:3:2$, we investigate the evolution of the admixtures over time, as shown in Fig.~\ref{Fig:CMSinitCondEvol}. The first observation is that at small times $t<1\rm fm$ the excited S-wave states feed the ground state and lead to its relative enhancement. This timescale aligns reasonably well with the rapid depopulation of the $\Upsilon'$ and $\Upsilon''$ states observed in Fig.~\ref{Fig:ExcitedBottom}. If we take a look at the evolution at later times, a replenishment of the excited states occurs with $\Upsilon '$ being the main benefactor as it is directly sourced by the ground state.

While it would be rather bold to claim any quantitative relation of this naive calculation with measurements in experiment it is reassuring that even in this highly idealized setting one observes at time $t<1 \rm fm$ a behavior akin to melting of the higher excited states, while populating the ground state. 

\subsection{Discussion}

The results for the evolution of a mixed initial state are encouraging in that they hint at excited state melting, obtained from a consistent potential-based description of heavy quarkonium including the effects of thermal decoherence while circumventing the interpretational difficulties associated with complex system energies. Nevertheless there is still a long way ahead of us before we can quantitatively compare to experiment or as would be the final goal to infer novel information about the medium from  the measured Bottomonium di-lepton spectra. 

\begin{figure}
 \includegraphics[angle=-90, scale=0.15]{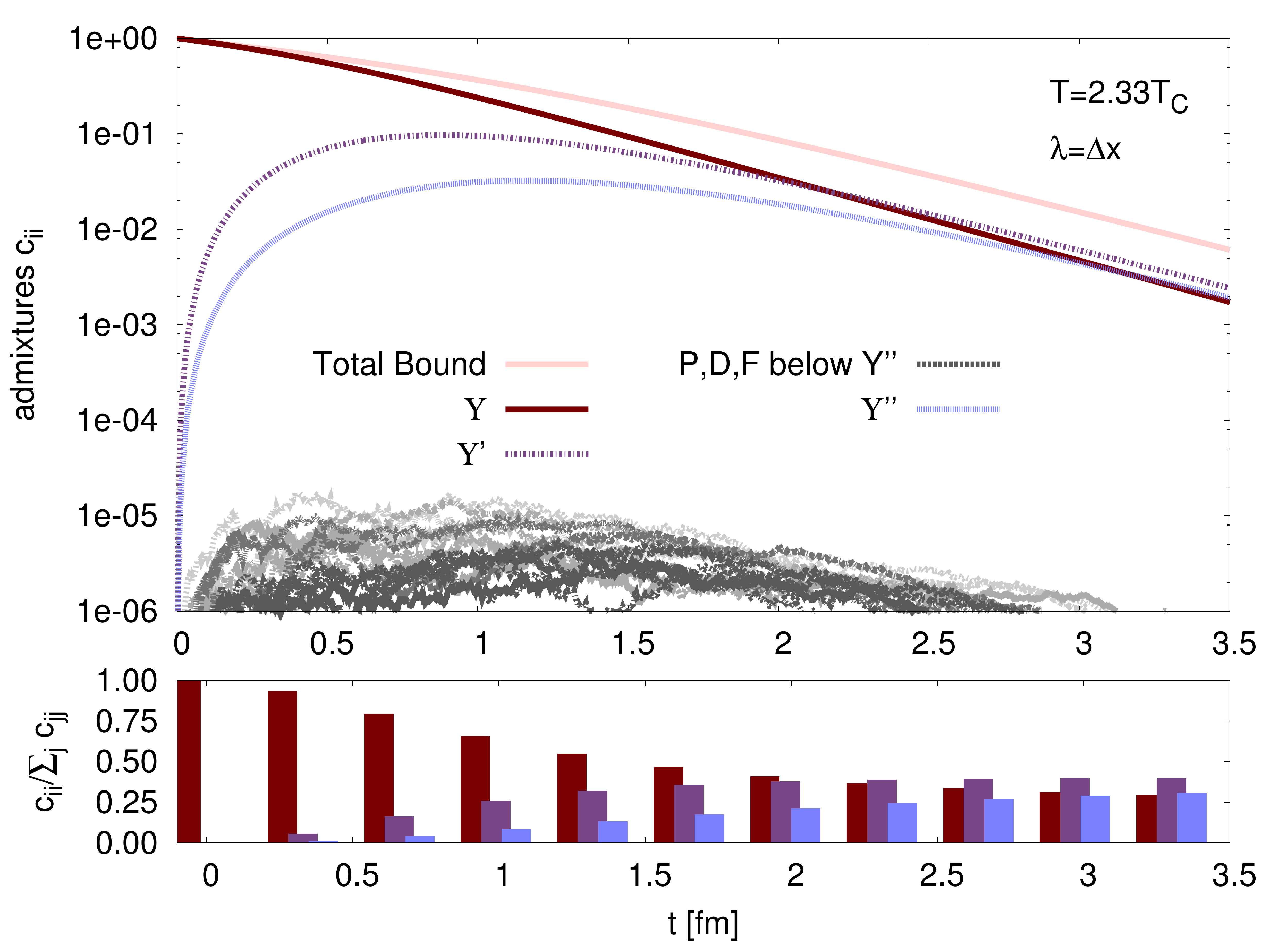}
 \includegraphics[angle=-90, scale=0.15]{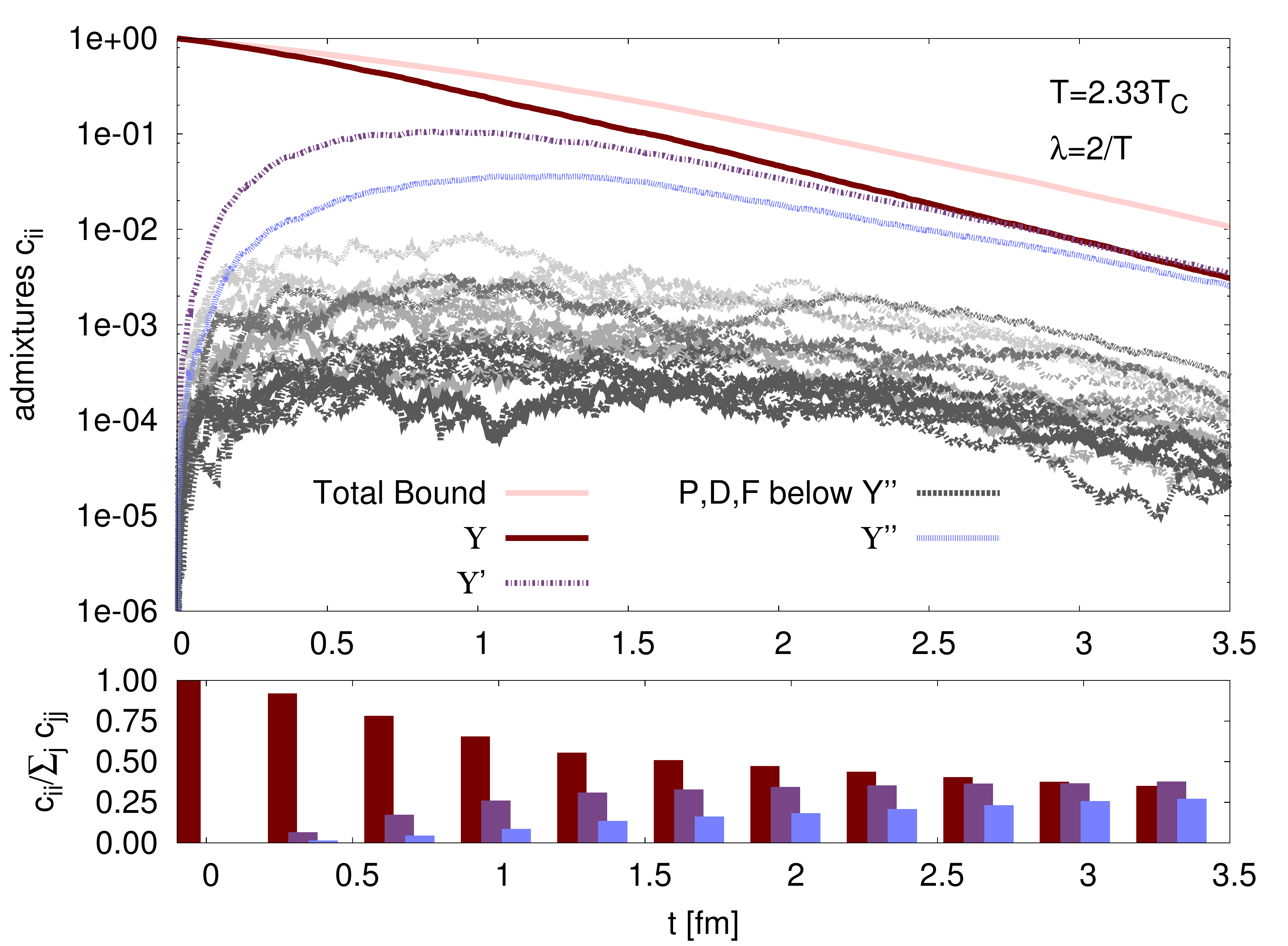}
 \caption{Evolution of a pure initial $\Upsilon$ state for two different choices of the noise correlation length. (left) $\lambda=0.03\rm fm\ll 1/T$: We find that a small correlation length weakens the ability of the noise for the excitation of states with different angular momentum. At the same time, the excitation of the S-wave states into the continuum is enhanced so that less bound states survive over time compared to $\lambda=1/T$ in Fig.~\ref{Fig:BottomStochAndNoNoise}. 
 (right) $\lambda=2\times 1/T\approx 1\rm fm$: The presence of fluctuations with relatively long correlations benefits the excitation of P and D states slightly compared to $\lambda=1/T$. Even though the effect is quite small the overall survival of the bound states is also found to be higher. }\label{Fig:DiffCorrLength}
\end{figure}

\begin{figure}
 \includegraphics[angle=-90, scale=0.15]{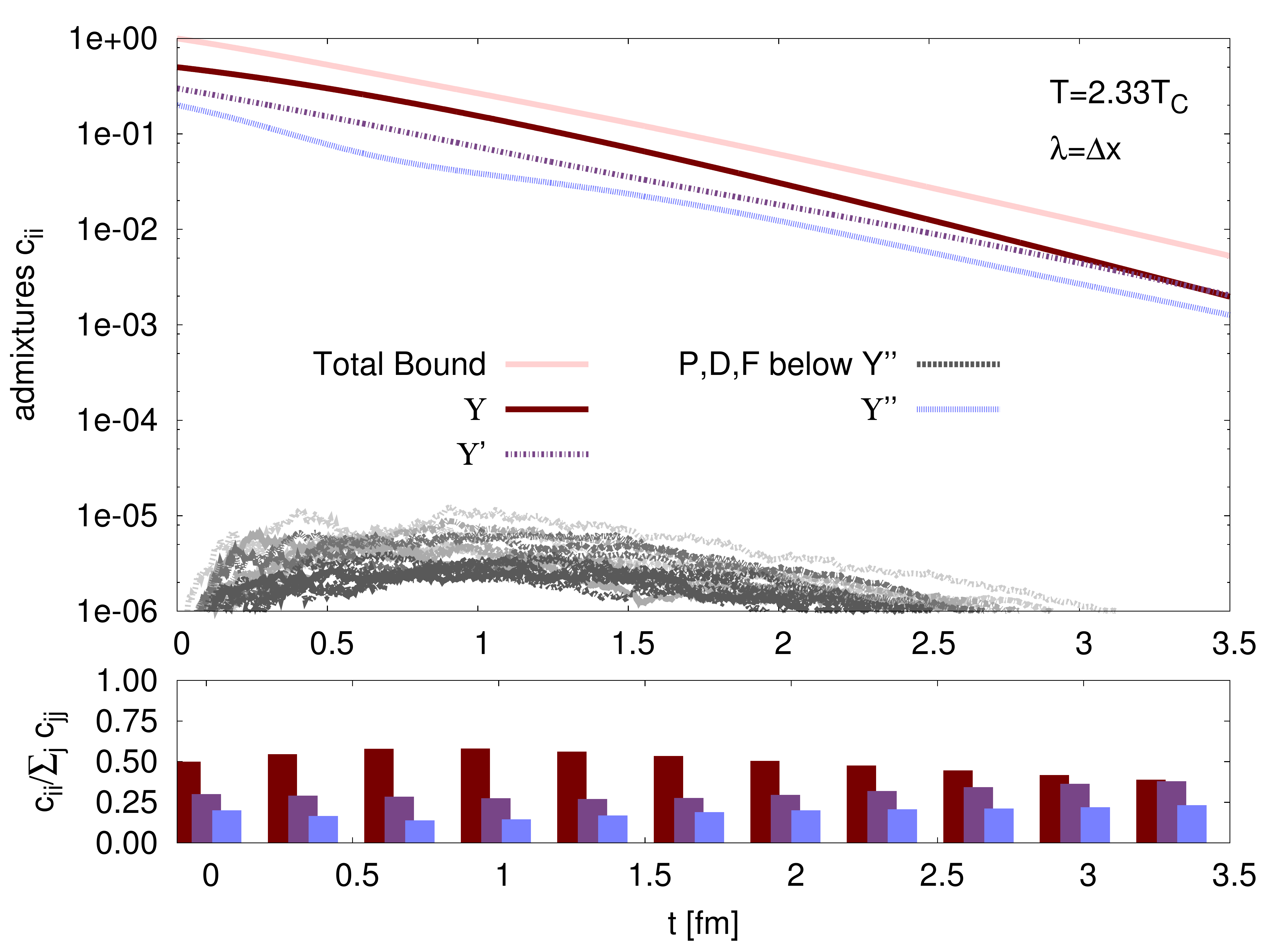}
 \includegraphics[angle=-90, scale=0.15]{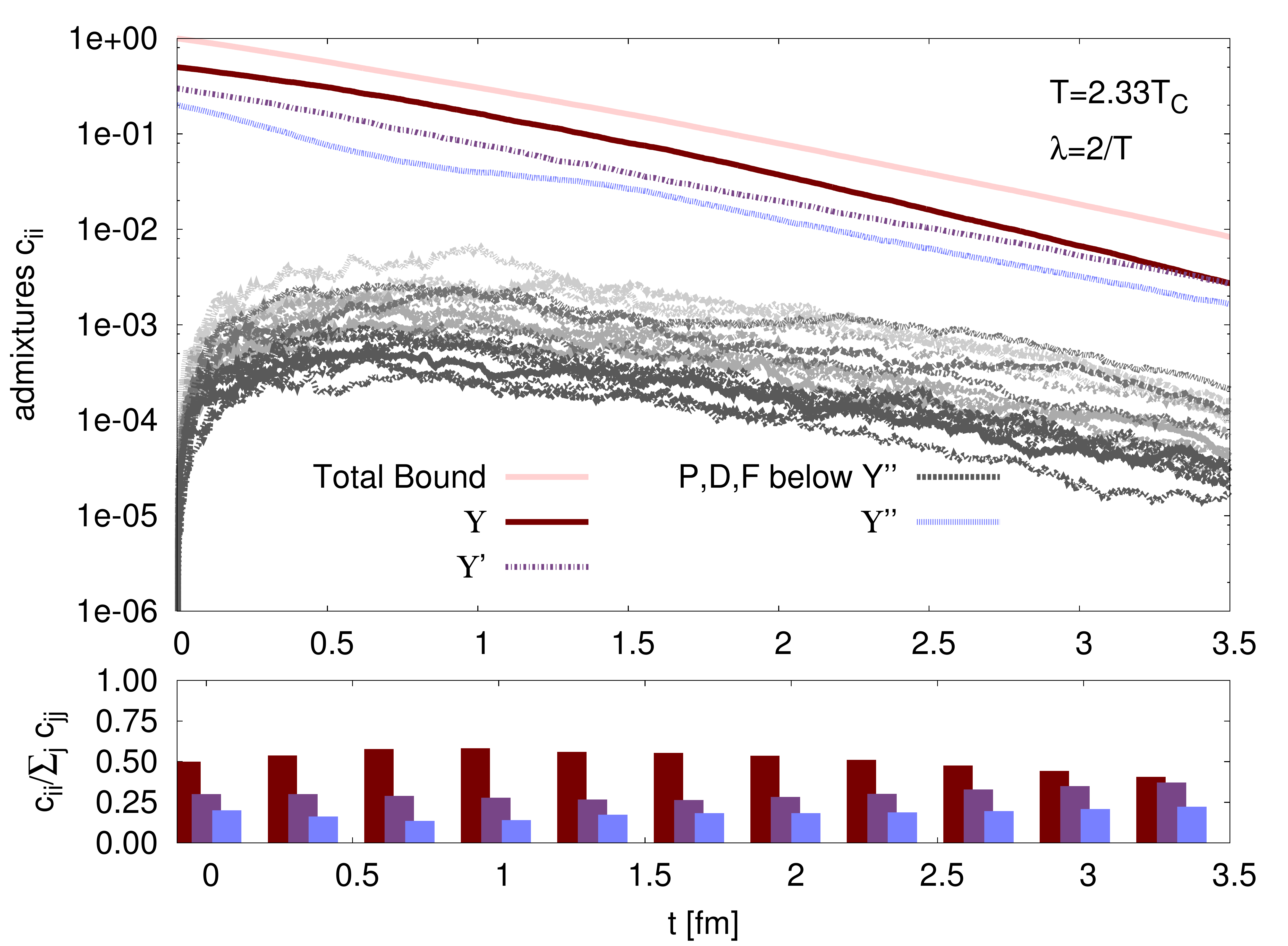}
 \caption{Evolution of the mixed initial state $\Upsilon:\Upsilon^\prime:\Upsilon^{\prime\prime}=5:3:2$ for two different choices of the noise correlation length. (left) $\lambda=0.03\rm fm\ll 1/T$: As in Fig.~\ref{Fig:DiffCorrLength}, a shorter correlation length weakens the ability of the noise to excite finite angular momentum states. Both at early and late times the relative abundances of the ground state are diminished relative to the excited states, as compared to the case of $\lambda=1/T$. 
 (right) $\lambda=2\times 1/T\approx 1\rm fm$: Long correlations benefit the excitation of P,D and F states. While at early times $t<2 \rm fm$ the relative contribution to the ground state survival is slightly diminished, at late times it is almost on par with the behavior found at $\lambda=1/T$.}\label{Fig:CMSDiffCorrLength}
\end{figure}

The first reason is that we do not yet have any reliable knowledge about the off-diagonal elements of the noise correlations. All the results shown above were based on a phenomenologically motivated choice of correlations, decaying according to a Gaussian with correlation length $\lambda=1/T$. At high temperatures with short correlation lengths such behavior might be plausible, close to the phase transition it is far from certain that it persists. Obviously we cannot ignore this issue as changing the values of $\lambda$, while not reflected on the level of the average wave-function or equivalently the thermal spectra, leads to discernible changes in the evolution of states shown in Fig.~\ref{Fig:DiffCorrLength} and Fig.~\ref{Fig:CMSDiffCorrLength} for the case of a pure and mixed initial conditions respectively. 

Starting with pure Bottomonium (Fig.~\ref{Fig:DiffCorrLength}), the presence of a small correlation length $\lambda=0.03\rm fm\ll 1/T$ diminishes the ability of the noise to excite states of different angular momentum, just as a larger value of $\lambda=2\times 1/T\approx 1\rm fm$ promotes their appearance. While the relative survival probability of the initial $\Upsilon$ at early times is not influenced strongly, beyond $t\sim 2 \rm fm$ shorter correlations lead to stronger ground state suppression and longer correlations favor its survival.

Interestingly, the interplay between the different states from a mixed initial condition, such as $\Upsilon:\Upsilon^\prime:\Upsilon^{\prime\prime}=5:3:2$ (Fig.~\ref{Fig:CMSDiffCorrLength}) for early times $t< 2 \rm fm$ shows consistently less relative enhancement of the ground state for both longer and shorter correlations than $\lambda=1/T$. At late times we still find however that long correlations slightly favor ground state survival, while shorter correlations diminish its contribution. 

These two examples show that the actual values of $\lambda$ and thus $\Gamma({\bf r},{\bf r}')$ are of phenomenological interest. It would thus be very helpful to find a field theoretical object which can serve as a counterpart in an EFT description. 
Just as the forward correlator allowed us to extract $V_{Q\bar{Q}}({\bf r})$ and $\Gamma({\bf r},{\bf r})$,  an appropriate four point meson correlator might supplement our knowledge on $\Gamma({\bf r},{\bf r}')$ for ${\bf r}\neq{\bf r}'$. 
 
The second and more fundamental reason is the inability for Eq.\eqref{Eq:StochSchroed} to correctly describe the thermalization of the heavy quarkonium with its surroundings. As was shown by Akamatsu in \cite{Akamatsu:2012vt}, if we wish to extend the predictive range of our computation to late times $t\sim10\rm fm$ we will have to make explicit the dissipative character of the dynamics by distinguishing forward and backward propagating modes and simulate them individually. 

The conceptually improved description \cite{Akamatsu:2012vt} takes into account the color degrees of freedom explicitly, reflected e.g. by the presence of colored noise, so that both the propagation of individual quarks and their bound states can be captured in medium. I.e. both color singlet and octet configurations are dynamical degrees of freedom. Inspection of their equation of motion hints at another limitation of the simple color neutral stochastic potential used here. As long as the heavy quarks in a singlet configuration remain close enough together, the gluons will not excite them into an octet state and we can safely assume as we did that the singlet propagates into a singlet. If however they become separated such that they appear as individual quarks to the medium, i.e. further than the thermal correlation length, it will quickly color rotate them into an octet configuration which then would have to be included explicitly.

The choice of initial conditions might also be ameliorated using the recently presented precision measurements for the $\chi_b$ states \cite{CMSPwave} in $p+p$ collisions. It will be interesting to see how these states with finite angular momentum evolve in the medium and whether their presence in a mixed ensemble does affect the S-wave abundances through (de)excitations induced by the stochastic noise.

Future simulations should also consider that the QGP actually evolves rapidly. It expands and cools, as can be explored by relativistic hydrodynamics simulations. While initially restricted to perfect liquids, fully causal approaches to viscous fluids are nowadays available \cite{Romatschke:2007mq,Akamatsu:2013wyk}. On the other hand progress has been made by constructing analytic solutions to the hydrodynamic equations \cite{Csanad:2012hr} that can take into account an equation of state provided by lattice QCD. After checking that the potential based approach remains valid in the presence of a rapidly changing temperature, the stochastic simulations should be amended by changing the parameters of the potential accordingly.

Another interesting feature to study within the stochastic approach is the momentum dependence of the heavy quarkonium melting process. At early times, i.e. before momenta are able to equilibrate, the $Q\bar{Q}$ state is certainly not at rest and it traverses the QGP. Studies of thermal spectral functions in lattice NRQCD \cite{Aarts:2012ka} as well as using EFT methods \cite{Escobedo:2013tca} revealed that its probability to remain in a well defined bound states is indeed affected by the velocity of the constituents. Since the potential based approach allows to equip the vacuum wave function with a finite momentum by a simple phase factor we should aim at simulating Bottomonium moving through an appropriately cooling medium. 

\begin{figure}
 \includegraphics[angle=-90, scale=0.15]{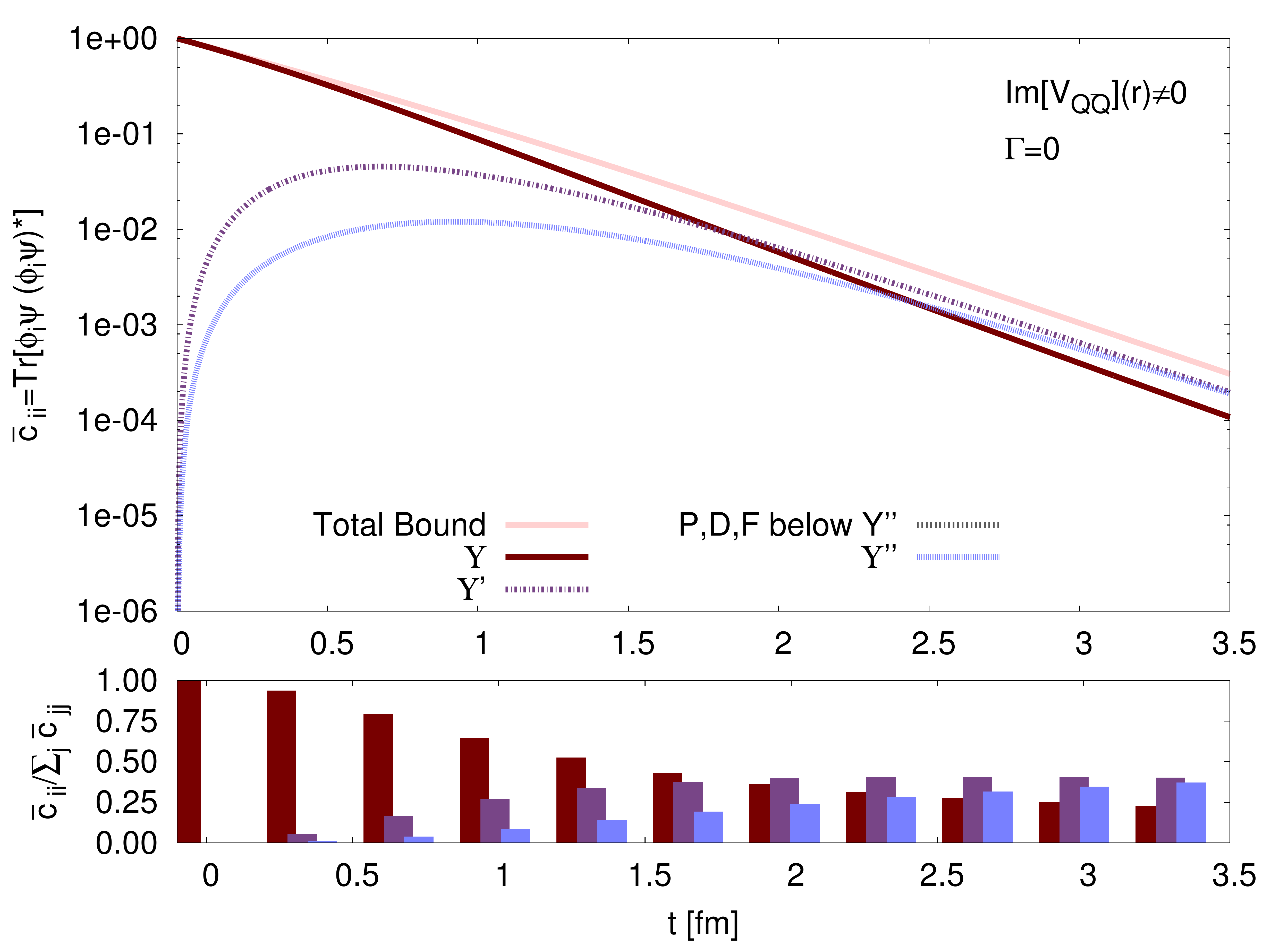}
 \includegraphics[angle=-90, scale=0.3]{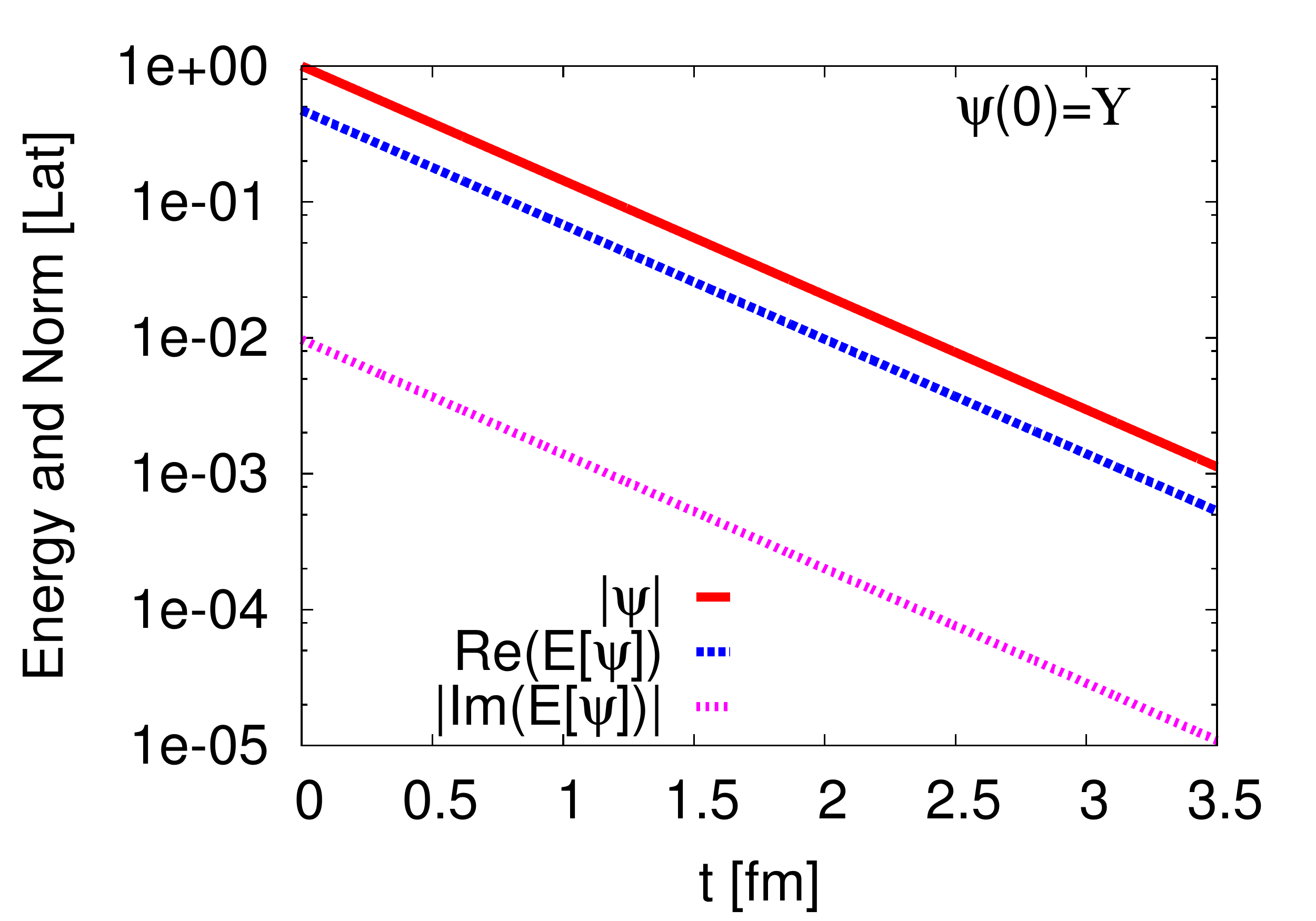}
 \caption{(left) Bottomonium evolution from an pure initial $\Upsilon$ state based on inserting an explicit imaginary part into $V_{Q\bar{Q}}({\bf r})$ in the Schroedinger equation \eqref{Eq:StochSchroed}, i.e. without stochastic noise $\Gamma({\bf r},{\bf r}')=0$ . As in the case for a purely real Hamiltonian we do not find any excitations of states with angular momentum different from zero. Comparing to the stochastic evolution in the right panel of Fig.~\ref{Fig:BottomStochAndNoNoise}, we find that e.g. the melting of the ground state appears to proceed more efficiently here. The damping of the wavefunction will ultimately lead to vanishing admixtures at late times and not to a thermal distribution. 
 (right) Decay of the wave function norm over time (red solid) and the accompanying decrease in the values of the system energy (blue).  Note that a finite but small imaginary part (magenta) is induced in the energies of the system and that the energy monotonously decreases. This reduction of the real-part of the energy does not comply with intuition. If a vacuum state is inserted into a thermal medium we would expect that the kicks of the medium constituents will actually transfer energy into the system until the a common temperature is reached eventually and the energy becomes time independent.}
 \label{Fig:ExplicitImagEvol}
\end{figure} 

\begin{figure}
\centering
 \includegraphics[angle=-90, scale=0.15]{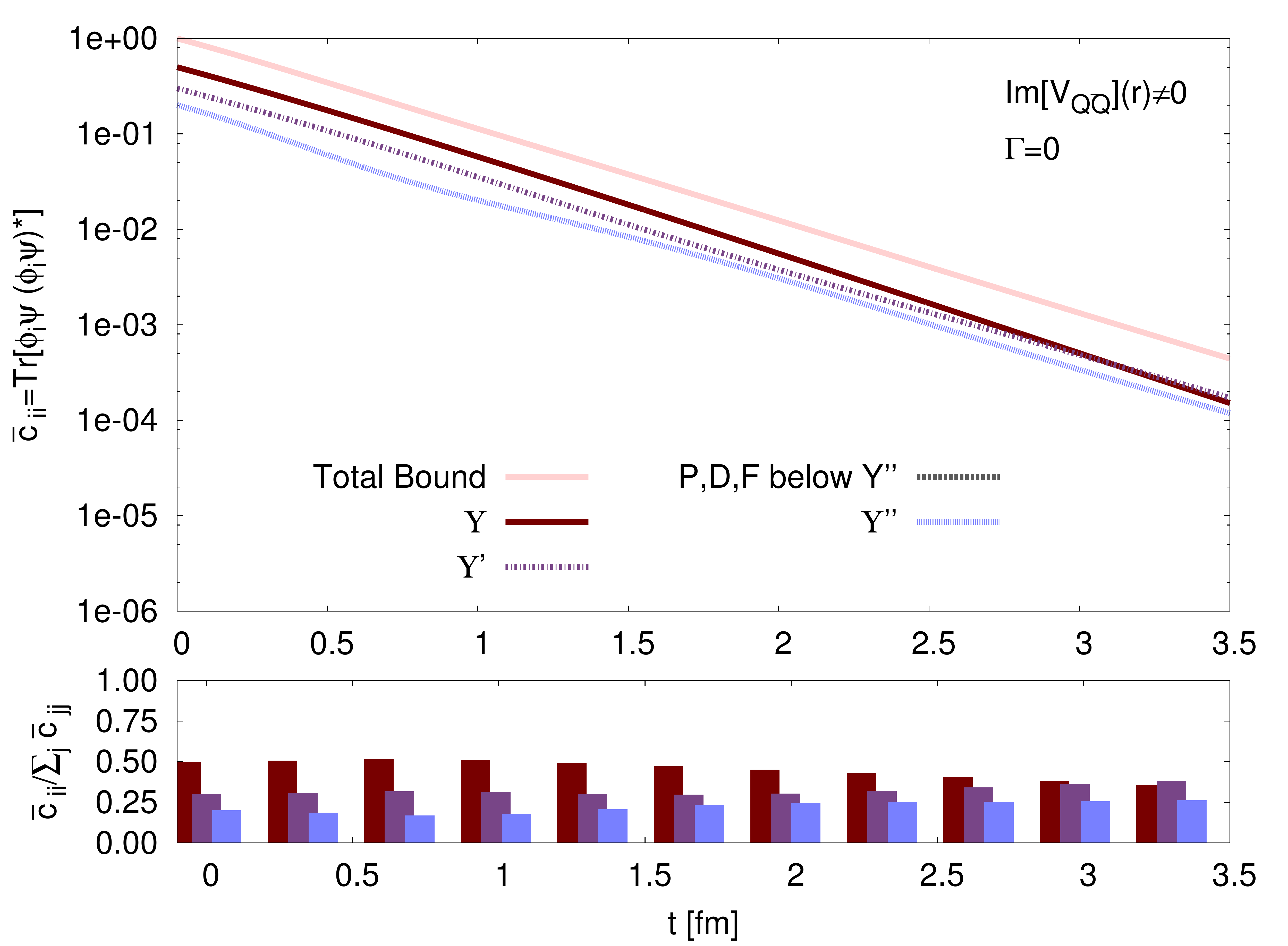}
 \caption{Evolution of the mixed initial state $\Upsilon:\Upsilon^\prime:\Upsilon^{\prime\prime}=5:3:2$  based on inserting an explicit imaginary part into $V_{Q\bar{Q}}({\bf r})$ in the Schroedinger equation \eqref{Eq:StochSchroed}, i.e. without stochastic noise $\Gamma({\bf r},{\bf r}')=0$ . As in the case for a purely real Hamiltonian we do not find any excitations of states with angular momentum different from zero. Comparing to the stochastic evolution, a clear difference in ground state survival emerges. Here at all times, the ground state is much less pronounced relative to the excited states, in contrast to the stochastic evolutions of either Fig.\ref{Fig:CMSinitCondEvol} or Fig.\ref{Fig:CMSDiffCorrLength}.}
 \label{Fig:CMSExplicitImagEvol}
\end{figure}

\section{Summary and Conclusion}

After revisiting the derivation of the Schroedinger equation \eqref{Eq:StatSchroed} and the associated effective field theory potential \eqref{Eq:DefPot} for the forward correlator of heavy quarkonium, we elaborated on the presence of an imaginary part of $V^{\rm EFT}({\bf r})$ in the QGP in terms of thermal decoherence. In order to consistently incorporate the effects of decoherence in a potential based, i.e. non-relativistic but otherwise dynamic, real-time description of Bottomonium wavefunctions (not correlators), we followed the open-quantum systems viewpoint introduced in \cite{Akamatsu:2011se}. 

Constructing an approximate unitary time evolution operator \eqref{Eq:StochTimeEvol} for the heavy quarkonium wavefunction $\psi_{Q\bar{Q}}({\bf r},t)$, based on a real but stochastic potential, we can calculate the admixtures $c_{nn}(t)$ of individual vacuum states $\phi_n({\bf r})$ as the system propagates in medium by appropriate projections \eqref{Eq:DefAdmix}. This approach invites the identification of the imaginary part of the EFT potential with the diagonal noise structure of the medium induced stochastic noise $\Gamma({\bf r},{\bf r})$, offering a (possibly non-perturbative) bridge between the EFT and the open-quantum systems language \cite{Akamatsu:2011se}. 
It differs thus from previous potential based studies (e.g.~\cite{Margotta:2011ta, Strickland:2011mw,CasalderreySolana:2012av}) in that no explicit non-hermiticity enters the heavy quarkonium Hamiltonian, all energies remain real-valued and actually increase over time (Fig.~\ref{Fig:PotentialsNrg} vs.~Fig.~\ref{Fig:ExplicitImagEvol}) as expected from inserting a vacuum state into the medium. In addition to these conceptual differences, the actual evolution of individual state admixtures is found to disagree quantitatively as well. Take e.g. the the case of mixed initial conditions (Fig.\ref{Fig:CMSExplicitImagEvol}), where ground state survival appears significantly lower than in the fully stochastic description.

The dependence of $c_{nn}(t)$ on the density matrix of states and thus on the off-diagonal elements of the noise correlations (see Fig.~\ref{Fig:DiffCorrLength}) entails that by observing Bottomonium states over time we can access information beyond what is encoded in the EFT potential, and hence the thermal spectral functions, which can be determined from $V^{EFT}({\bf r})$. I.e. if we already knew the temperature of the medium created in relativistic heavy ion collisions and a reliable EFT potential is used, the measured abundances of Bottomonium might contribute to shedding light on the correlation structure of thermal fluctuations within the QGP.

As a first attempt at estimating the behavior of different Bottomonium vacuum states as they propagate through the QGP, we investigate a highly idealized static thermal medium. At a temperatures of $T=2.33T_C$ the perturbative EFT potential by Laine is used to set the parameters $V_{Q\bar{Q}}({\bf r})$ and $\Gamma({\bf r},{\bf r})$, while for the off-diagonal entries we make a simple Gaussian ansatz with a characteristic correlation length of $\lambda=1/T$ (Fig.~\ref{Fig:PotentialsNrg} and Fig.~\ref{Fig:NoiseCorrRealiz}).

As was known from the study of model potentials and has been emphasized recently in \cite{Borghini:2011yq,Borghini:2011ms, Akamatsu:2011se, CasalderreySolana:2012av, Dutta:2012nw}, we find that the mixing of different vacuum Eigenstates due to the presence of a screened real part of the heavy quark potential $V_{Q\bar{Q}}({\bf r})$ is a significant contribution to the overall dynamics (Fig.~\ref{Fig:BottomStochAndNoNoise}). Note that in the absence of thermal noise (i.e. also for purely real EFT potentials), only states with the same angular momentum as the initial state participate, which however also includes unbound continuum states.

In the presence of thermal fluctuations, i.e. when taking into account the imaginary part of the EFT potential, the admixture of bound states decreases much more rapidly. Starting from pure initial states (Fig.~\ref{Fig:BottomStochAndNoNoise} and Fig.~\ref{Fig:ExcitedBottom}), one finds that an intricate interplay between mixing and the stochastic (de)excitation of states leads to non-trivial patterns of decay and replenishment of the S-wave states. Since the noise can also excite states of different angular momentum, we observe a finite but relatively small admixture of $P,D,F,...$ states.  Until the open-heavy flavor measurements currently performed at the LHC have established a baseline for the overall number of $b\bar{b}$ produced in $\rm Pb+Pb$ collisions we refrain from absolute comparisons and focus on the relative abundances of the individual states. 

Starting from a mixed initial state consisting of $\Upsilon:\Upsilon':\Upsilon" = 5:3:2$ ratio of S-wave states, inspired by the dilepton spectra determined by the CMS collaboration in $\rm p+p$ collisions, a mild overall melting effect of the excited states is obtained at early times $t<1\rm fm$ (Fig.~\ref{Fig:CMSinitCondEvol}). We find a tendency for regeneration of the excited states at later $t>1.5\rm fm$ which is to  be expected from the behavior of the pure initial state runs. Due to the very naive setting of Bottomonium at rest in a static thermal medium this result should not be thought of as a quantitative statement but rather as a motivation to develop further the open-quantum systems approach to heavy quarkonium in general and the stochastic potential framework in particular, aiming at more realistic scenarios.

The first step will be to include the effects of an evolving temperature obtained from numerical or analytic solutions of the hydrodynamic equations. Together with endowing the initial wavefunction with a finite momentum and taking into account more than just the initial S-wave admixtures it should be possible to obtain a more realistic estimation of the Bottomonium evolution at early times. The implementation of the improved formulation of the stochastic dynamics based on the Feynman-Vernon influence functional should concurrently be pursued to enable a more reliable investigation of the late time approach to thermal equilibrium which we expect will be of use in building a bridge from the language of thermal lattice QCD spectral functions to the phenomenology of Bottomonium melting.

The author thanks Y. Akamatsu and Y. Burnier for stimulating discussions, as well as for carefully reading the manuscript. This work was partly supported by the Swiss National Science Foundation (SNF) under grant 200021-140234.

\end{document}